\documentclass[titlepage]{article}

\pdfoutput=1
\usepackage{arxiv}

\usepackage{setspace}
\usepackage[section]{placeins}

\usepackage[authoryear,round]{natbib}

\usepackage[utf8]{inputenc} 
\usepackage[T1]{fontenc}    
\usepackage[hidelinks]{hyperref}       
\usepackage{url}            
\usepackage{booktabs}       
\usepackage{amsfonts}       
\usepackage{nicefrac}       
\usepackage{microtype}      
\usepackage{lipsum}
\usepackage{graphicx}
\usepackage{amsmath}
\usepackage{amsfonts}
\usepackage{amssymb}
\usepackage{multirow}
\usepackage{xcolor}
\usepackage{array} 
\graphicspath{ {./images/} }
\usepackage{algorithm}
\usepackage{algorithmic}

\usepackage[most]{tcolorbox}
\usepackage{blindtext}
\usepackage{lineno}
\usepackage{subcaption}

\makeatletter
\let\@fnsymbol\@arabic
\makeatother

\title{The ecological forecast limit revisited: Potential, actual and relative system predictability}

\author{Marieke Wesselkamp \thanks{Department of Biometry, University of Freiburg, Germany} \\ \And Jakob Albrecht \thanks{Department of Forest Economics, University of Freiburg, Germany} \\ \And Ewan Pinnington \thanks{European Center for Medium-Range Weather Forecasts, United Kingdom} \\ \And William J. Castillo \footnotemark[1] \\ \And Florian Pappenberger \footnotemark[3] \\ \And Carsten F.~Dormann\footnotemark[1]}

\begin{document}

\maketitle

\begin{abstract}

Ecological forecasts are model-based statements about currently unknown ecosystem states in time or space. For a model forecast to be useful to inform decision makers, model validation and verification determine adequateness. The measure of forecast goodness that can be translated into a limit up to which a forecast is acceptable is known as the `forecast limit'. While verification in weather forecasting follows strict criteria with established metrics and forecast limits, assessments of ecological forecasting models still remain experiment-specific, and forecast limits are rarely reported. As such, users of ecological forecasts remain uninformed of how far into the future statements can be trusted. In this work, we synthesise existing approaches to define empirical forecast limits in a unified framework for assessing ecological predictability and offer recipes for their computation. We distinguish the model's potential and absolute forecast limit, and show how a benchmark model can help determine its relative forecast limit. The approaches are demonstrated with three case studies from population, ecosystem, and Earth system research. We found that forecast limits can be computed with three requirements: A verification reference, a scoring function, and a predictive error tolerance. Within our framework, forecast limits are defined for practically any ecological forecast and support research on ecological predictability analysis. 

\end{abstract}

\section{Introduction}

Interest in and use of forecasting models for understanding and managing ecosystems has steadily increased in recent years \citep{record_synthesizing_2023, dietze_near-term_2024}. The relevance of forecasting Earth and environmental systems is driven in part by two large-scale changes: On the one hand, the warming of the global climate affects ecosystems of all types. On the other hand, the shift in modelling paradigm across various disciplines towards techniques generally summarised under `artificial intelligence', which in Earth system modelling most notably happened in the field of weather forecasting \citep{keisler_forecasting_2022, bauer_deep_2023, ben_bouallegue_rise_2024} and hydrology \citep{zwart_nearterm_2023, nearing_global_2024}, but also in ecosystem ecology and land surface forecasting \citep{getz_making_2018, wesselkamp_advances_2025}. 
Forecasts that fall into the realm of ecosystem ecology range from population dynamics \citep[e.g.][]{daugaard_forecasting_2022, karunarathna_modelling_2024}, plant phenology \citep[e.g.][]{wheeler_predicting_2024}, to predicting forest productivity on site \citep{kazimirovic_dynamic_2024} and landscape scales \citep{seidl_individual-based_2012}, and also encompass near-term lake \citep{thomas_nearterm_2020} or streamflow \citep{zwart_nearterm_2023} temperature forecasting. The Ecological Forecasting Initiative (EFI) is hosting the community-wide NEON (National Ecological Observatory Network) forecasting challenge, which accelerates scientific learning across ecological themes such as terrestrial water and carbon fluxes, tick populations, or plant phenology \citep{thomas_span_2023}. To analyse results from diverse forecasting studies and to improve communication with decision-makers on climate change action \citep{dietze_near-term_2024, wheeler_predicting_2024}, it is essential to adopt best practices and to understand the predictability of ecological systems regarding their properties and forecast horizons \citep[][, see Box 1 for definition]{petchey_ecological_2015, lewis_power_2022}. 


Ecological forecasts have been defined as ``the process of predicting the state of ecosystems [...] with fully specified uncertainties, [...] contingent on explicit scenarios'' of environmental conditions \citep[][p.657]{clark_ecological_2001}. Further, (1) they are temporal predictions \citep{dietze_ecological_2017}, (2) either aim to inform the general public \citep{thomas_nearterm_2020, schaeffer_forecasting_2024} or function as standalone tools for assisting stakeholders and scientific learning \citep{urban_coding_2022, dietze_near-term_2024}, and (3), they are near-term on daily to decadal time scales and can be updated with new observations \citep{dietze_ecological_2017}. Both scientific learning and making decision-relevant forecasts will benefit from a deeper understanding of the predictability of ecological systems and variables \citep{dietze_near-term_2024}. One approach to measure system predictability is the temporal forecast limit, which can inform conservation strategies, such as identifying the time frames within which species population models remain reliable to predict biodiversity outcomes in different climate scenarios \citep{petchey_ecological_2015, urban_coding_2022}. This is particularly important in adaptive management frameworks, where the timing and duration of interventions can depend on the predictability of ecosystem responses \cite{dietze_near-term_2024}, and quantifying the forecast limit will support anticipating implications for management decisions \citep{clark_ecological_2001, petchey_ecological_2015}. By establishing a formalised approach to forecast limits, we aim to provide a standardised tool that assists ecological forecasters in assessing the temporal reliability of their models. 

The concept of forecast limits in ecology was introduced ten years ago as the ``forecast horizon'' \citep{petchey_ecological_2015}. It is a response to the discovery of chaos in non-periodic deterministic flows of atmospheric dynamics \citep{lorenz_deterministic_1963, hunt_theory_2004}, stating that two trajectories of an unstable system with slightly different initial states have finite predictability. Today, forecast limits are established in meteorology \citep{buizza_forecast_2015, magnusson_dependence_2019}, in hydrology \citep{reggiani_time-horizons_2024}, and since the work of \citet{petchey_ecological_2015} also in ecology \citep{adler_matching_2020, woelmer_nearterm_2022}.
The forecast horizon refers to the time period over which future projections and predictions are made (see Box 1). Over this horizon, the forecast limit defines the moment when the forecast quality is not better than a reference model or, if available, a pre-defined standard \citep{petchey_ecological_2015, dietze_near-term_2024}. Ideally, a forecast is reported up to its forecast limit or at least indicates such a limit \citep{dietze_near-term_2024}; to our knowledge, the terminology is not consistently defined in the literature, and we refer to Box 1 for the definitions used in this work.
This defines the forecast limit based on forecast quality, assessed through scoring metrics of forecast performance. Scoring metrics measure the dissimilarity of the forecast distribution and the observation; smaller is better, and they rely on verification data \citep{murphy_what_1993}. The definition of a forecast limit is inversely related to the concept of system predictability, which is the theoretical study of predictive ability under uncertainty (see also Box 1) \citep{palmer_predictability_2006}. 
However, forecast performance alone does not necessarily equate to forecast utility, which considers how forecasts inform decision-making in practical contexts from a user perspective. Forecast utility depends on whether the predictive error remains within an acceptable range for a given application \citep{murphy_what_1993}. For example, a weather forecast with a 2°C error margin might be sufficient for general use but could prove inadequate for agricultural decisions, where precision is critical for crop survival. Therefore, determining a forecast limit using performance measures represents a technical perspective \citep{adler_matching_2020, thomas_nearterm_2020, woelmer_nearterm_2022}, but the practical utility depends on the needs and risk tolerance of forecast users.

Against this background, we propose determining predictability with forecast limits in a framework that applies to point forecasts, i.e.~single forecast trajectories, and to probabilistic forecasts, i.e.~forecast distributions. Based on the above definition, this requires scoring metrics that quantify the magnitude of predictive error either deterministically or probabilistically \citep{dalcher_error_1987, murphy_what_1993}. This framework is based on two assumptions: First, predictability is inversely related to forecast uncertainty, hence the forecast limit decreases as forecast uncertainty increases \citep{palmer_predictability_2006}. Second, a deterministic evaluation can be appropriate when predictive error and forecast uncertainty are correlated \citep{hopson_assessing_2014}. Verifying observations for forecast evaluation over the forecast period needs to be available to determine realisable predictability, i.e.~model predictive ability (see Figure \ref{fig:timeline}B, Label 1) \citep{pennekamp_intrinsic_2019}. Without a verification, model-intrinsic predictability can be determined (see Figure \ref{fig:timeline}B, Label 2). An additional requirement is the explicit statement of a tolerance for the predictive error up to which forecast performance is acceptable \citep{petchey_ecological_2015, owens_ecmwf_2018, mcwilliams_perspective_2018}. Such tolerance may come as an \textit{ad-hoc} expectation towards the score, but more commonly it is a \textit{benchmark} or \textit{null model} for which performance can be determined equivalently to the forecast model (see Figure \ref{fig:timeline}B, Label 3) \citep{buizza_forecast_2015, massoud_probing_2018, thomas_nearterm_2020}.

\begin{tcolorbox}[breakable, title = Box 1. Terminology.]

\begin{figure}[H]
    \centering
    \includegraphics[width=0.99\linewidth]{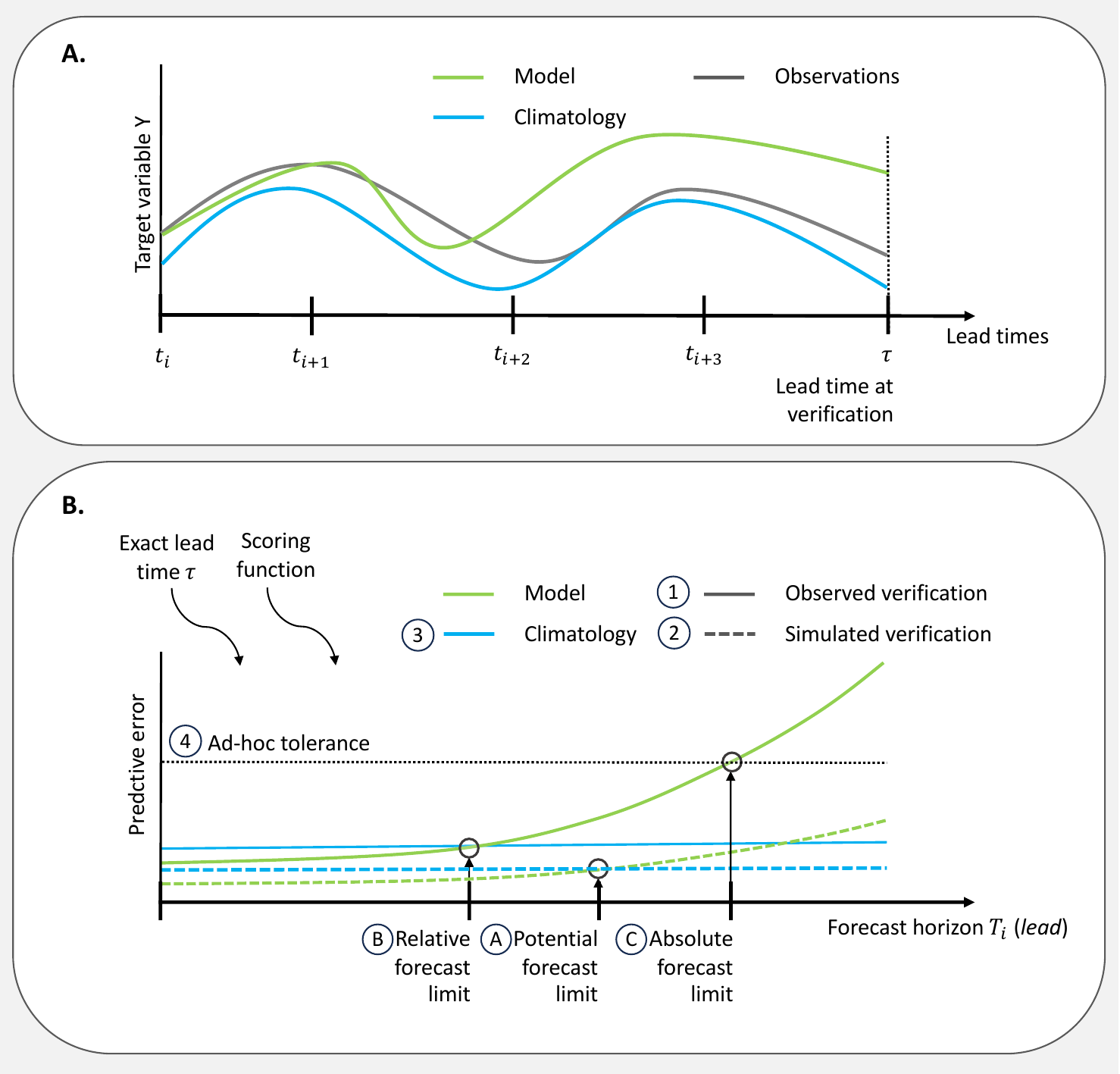}
    \caption{Terminology in determining temporal forecast limits. \textbf{A.} Single forecast of a target variable $Y$ as a function of lead time $\tau$. \textbf{B.} The predictive error at an exemplary verification time $\tau$ as a function of the lead, i.e.~the forecast horizon $T_i$. The climatology (3) (or: long-term mean) is the expected value at $\tau$. The forecast is made from an arbitrary initial forecast time $t_i$ up to $\tau$ over the horizon $T_i$. The relative forecast limit (B) is estimated from predictive error by evaluation with observations (1), relative to the reference model, i.e.~the climatology. The potential forecast limit (A) is estimated by evaluating the forecast and climatology against simulations (2). The absolute forecast limit (C) is estimated with an ad-hoc tolerance (4) towards the predictive error.}
    \label{fig:timeline}
\end{figure}

The terminology for different terminological concepts in temporal forecasting has, to our knowledge, not been consistently used across disciplines. Being aware of other definitions, for the course of this work, we will use the vocabulary as introduced below.

\paragraph{Initial forecast time $\mathbf{t_i}$} The (arbitrary) moment in time where a forecast is initialised.

\paragraph{Lead time $\mathbf{\tau}$} The exact date-time for which the forecast is made. 

\paragraph{Forecast horizon $\mathbf{T_i}$} The period of time over which a forecast is made from $t_i$, i.e.~the `lead' in lead time.

\paragraph{Forecast limit $\mathbf{h}$} Moment in time at which predictive error, measured with a scoring function $\mathcal{S}$, exceeds $\varrho$, which is a reference model or an ad-hoc value \citep{petchey_ecological_2015, buizza_forecast_2015, dietze_near-term_2024}. It is also known as predictability limit \citep{spring_predictability_2020} and, in conflict with meteorological definitions of that term, has previously been termed forecast horizon \citep{petchey_ecological_2015}. 

\paragraph{Verification time} Moment in time at which forecast is evaluated against an observation.

\paragraph{Climatology} The long-term expectation of a cyclic system variable. It is described by a distribution of past observations from which a climatological (originally for weather-related variables) mean and standard deviation can be derived and can be used as a benchmark model, as exemplified in Figure \ref{fig:timeline}B (Label 3).

\paragraph{Predictability} The viewpoint that the ability to predict the future state of a system is limited by forecast uncertainty \citep{boffetta_predictability_2002}. It is under certain conditions correlated with predictive ability \citep{palmer_predictability_2006, pennekamp_intrinsic_2019}, or equivalently used \citep{shen_lorenzs_2023} and also known as intrinsic or potential predictability \citep{tiedje_potential_2012, sun_intrinsic_2016, pennekamp_intrinsic_2019}.

\paragraph{Predictive ability} The quantified skill to predict the future states of a system, based on a scoring function and an observational verification. It is also known as realised or practical predictability \citep{smith_uncertainty_1999, pennekamp_intrinsic_2019}.

\end{tcolorbox}

Within this framework, different forecast limits can be computed (see Figure \ref{fig:timeline}). When verification data is available for one or more lead times (see Figure \ref{fig:timeline}, Label 1), the realised forecast performance can be determined over lead times. Typically, a reference model, that is benchmark or null model (see Figure \ref{fig:timeline}B, Label 3), sets the scoring tolerance by simple comparison \citep[e.g.][]{wheeler_predicting_2024}. However, in contrast to just plotting forecast and reference model performances over lead times, skill scores assess forecast and reference model performances relatively \citep{pappenberger_how_2015} and indicate the \textit{relative forecast limit} as the moment in time when the forecast model is no better than the reference model (see Figure \ref{fig:timeline}B, Label 4). For example, a forecast ensemble of the European Centre for Medium-range Weather Forecasts (ECMWF) of geopotential height that was computed with the climatological ensemble as reference model, which is the historical model forecast ensemble constrained by observations, referred to this forecast limit as the ``forecast skill horizon'' \citep{buizza_forecast_2015}.
Model-intrinsic predictability can be studied without observational verification data and be quantified with the \textit{potential forecast limit} (see Figure \ref{fig:timeline}B, Label A), as has been done repeatedly in seasonal forecasting, for example for the Atlantic Meridional Overturning Circulation \citep{msadek_assessing_2010}, meridional heat transport \citep{tiedje_potential_2012}, and for net primary productivity in marine ecosystems \citep{frolicher_potential_2020, buchovecky_potential_2023}. Here, forecast uncertainties are controlled by assuming perfect system knowledge and evaluating forecast performance against the forecast ensemble mean or simulated forecast ensemble members (see Figure \ref{fig:timeline}B, Label 2) \citep{seferian_assessing_2018, spring_predictability_2020, delsole_statistical_2022}. Finally, an estimate of the \textit{absolute forecast limit} is the point in time when forecast performance exceeds an ad-hoc defined tolerance to the score (see Figure \ref{fig:timeline}B, Label 6)  \citep{petchey_ecological_2015}. While such tolerance is not often available ad-hoc, the effect of this choice is comprehensibly demonstrated in \citet[e.g.][]{massoud_probing_2018} with a plankton community model, displaying the forecast limit as a function of ad-hoc tolerance.

The goal of this work is to synthesise existing methods that assess the above-introduced forecast limits and apply them to quantify model predictability and system behaviour. This work explores the relationship between ecological model verification and predictability analysis. It formalises the distinctions between the potential and relative forecast limit that we assume to represent the upper and lower predictability limits, and a use case of the absolute forecast limit. We demonstrate this framework using three case studies that address three different ecological scales: 
\begin{itemize}
    \item The potential forecast limit using a single-species population model (stochastic Ricker equation).
    \item The absolute forecast limit using iLand, an individual-based forest model \citep{seidl_individual-based_2012}.
    \item The relative forecast limit using aiLand, a machine learning emulator for land-surface modelling \citep{boussetta_ecland_2021, wesselkamp_advances_2025}.
\end{itemize}
These case studies exemplify the diverse applications of forecast limits in evaluating models across different systems and complexities.

\section{Material and Methods}

Our formalisation heavily leans on \citet{dietze_prediction_2017} and \citet{berger_statistical_2019}. The notation follows a statistical tradition: model statements are indicated as $\widehat{Y}$, interpreted as estimates of the system states, the closest approximation to which are observed verifications $Y$. This section contains the general and shared methods. Because our case studies strongly differ in modelling approaches, experiment-specific methodological decisions that follow from the general methods will be given in an experimental setup description for each case study for convenience, along with the results.

\subsection{Models, forecasts and their errors}\label{sec:forecast_models}

We define a forecast model, $\mathcal{M}(Y_0, X, \theta)$, as a function or family of functions of initial state variables $Y_0$, optional external forcing variables $X$, and model parameters $\theta$ \citep{dietze_prediction_2017}. 
The model $\mathcal{M}$ at each time step, $t$, returns an estimate, $\widehat{Y}_t$, of the system state. $\mathcal{M}$ can be initialised with $Y_0$ at an arbitrary date, the initial forecast time $t_i$. The date to which $\mathcal{M}$ forecasts is the lead time $\tau$. The period of time from $t_i$ over which this is happening, i.e.~the lead, $T_i = \{ t_i, t_{i+1} , \dots, \tau\}$, is the forecast horizon. 
Since there will always be some discrepancy between the predicted state of the system and its true state, $\widehat{Y}_t$ incurs process error $\eta_t$. This process does not represent observation error, but model structural error: the dynamical uncertainty in the predicted transition between time steps  \citep{berger_statistical_2019},
\begin{equation}\label{eq:forecast_model}
\widehat{Y}_t = \mathcal{M}(\widehat{Y}_{t-1}, X, \theta) + \eta_t \quad\text{with}\quad t \in T_i \ ,
\end{equation}


The predictive or forecast error $\varepsilon_t$ defines the difference between a single estimate of the model, at time $t$, with the best approximation to the ``true'' process value, the verification $Y_t$ 
\citep{judd_geometry_2008} as: 
\begin{equation}\label{eq:predictive_error}
    \varepsilon_t = \widehat{Y_t} - Y_t\ .
\end{equation}


Based on the introduced uncertainties in any of the components of equation 
\ref{eq:forecast_model} \citep{dietze_ecological_2017}, 
consider an ensemble of independent estimations of the process; 
Then, select the realisations corresponding to time step $t$: 

\begin{equation}\label{eq:forecast_ensemble}
{ \{ \widehat{ Y }_t^{M} \} }_{t \in T_i}  =  \left\{ \widehat{ Y }_t^{(1)}, \widehat{ Y }_t^{(2)}, \dots, \widehat{ Y }_t^{(M)} \right\}_{t \in T_i} \ ,
\end{equation}

This can be described as a sample of $M$ realisations of model forecasts $\widehat{Y}_t$, which follow the probability density function (PDF) $p_{Y_t}({\widehat{Y}_t})$. 
Then, at each time step $t$, when comparing the verification with the ensemble of estimates (as given in Equation \ref{eq:forecast_ensemble}), the PDF of the predictive error has the same functional form as $ { \{ \widehat{ Y }_t^{(m)} \}}_{m = 1}^{M} $, i.e.:
\begin{equation}\label{eq:probabilistic_error}
    \{ \varepsilon_{t}^{M} \} \sim p_{Y_t}({\widehat{Y}_t})
\end{equation}

Identities \ref{eq:predictive_error} and \ref{eq:probabilistic_error} describe how we evaluate a model's performance and predictability, respectively. Although these quantities have different dimensions, the following general definitions apply to both cases. 

\subsection{The forecast limit}\label{sec:formalisation}
Following the definition of \citet{petchey_ecological_2015} and \citet{buizza_forecast_2015}, we formalise the forecast limit, $h$, as a function of $\tau$ (see Box 1, Figure \ref{fig:timeline}A). 
Computing  $h$ requires a scoring function to assess forecast quality and a tolerance towards the returned score (Figure~\ref{fig:timeline} B, Label 4). 
We define a scoring function $\mathcal{S}$ \citep{gneiting_strictly_2007} as any monotonic function that evaluates the correspondence of either $\widehat{Y}_{t}$ and $Y_{t}$ for deterministic forecasts (Eq. \ref{eq:predictive_error}) or of $\{ \widehat{ Y }_t^{M} \}$ and $Y_{t}$ for probabilistic forecast (Eq. \ref{eq:probabilistic_error}). When $\mathcal{S}$ evaluates Eq. \ref{eq:predictive_error} or the full distribution in Eq. \ref{eq:probabilistic_error} at once, $h$ is a scalar. If it evaluates Eq. \ref{eq:probabilistic_error} pointwise, $h$ has a PDF itself.  
We additionally define the scoring tolerance $\varrho = \varrho( k ) $, that can be a function of time or space, or any other (environmental) variable $k$ as a threshold applied to $\mathcal{S}$ that establishes the predictive error tolerance, i.e.: 
\begin{equation}
    \label{eq:ScoringInequation}
    \mathcal{S}(\varepsilon_t) \leq \varrho \ \quad\text{with}\quad t \in T_i \ .
\end{equation}
Then, the first point in time at which the last condition no longer holds ($ \mathcal{S}(\varepsilon_t) = \varrho $) is the estimated forecast limit $h$
\begin{equation}\label{eq:horizons}
h = \underset{t}{\text{argmin}} \left(  \mathcal{S}(\varepsilon_t) - \varrho  \right) \quad\text{where}\quad  t \in T_i \, 
\end{equation}

and the evaluation of the expression in the brackets of \ref{eq:horizons} at each time step $t$ can be expressed as a step function:
\begin{equation}
\label{eq:step_function}
\gamma = 
\begin{cases} 
1 & \text{if } \mathcal{S}(\varepsilon_t) - \varrho = 0 \\
0 & \text{otherwise}
\end{cases}
\quad\text{where}\quad  t \in T_i .
\end{equation}
Thus, equation \ref{eq:step_function} provides the estimate for $h$ when the step function $\gamma$ changes from 0 to 1. 
When confronted with an observational verification, $h$ estimates the model's realisable predictability, which can be seen as the lower bound of a model's forecast limit \citep{lorenz_atmospheric_1982, pennekamp_intrinsic_2019}.


\subsection{Relative skill}\label{sec:relative_skill}

Commonly, a benchmark or low-skill null model serves as a reference and, by condition \ref{eq:ScoringInequation}, defines $\varrho$ as a function of time, $\varrho(t)$ \citep{jolliffe_forecast_2003, pappenberger_how_2015}. 
Such reference model, $\mathcal{R}(Y_0, X, \theta)$, may take several forms, including the long-term mean at an exact lead time, $\mathcal{R}(Y_{\tau}) = \mathbb{E}(Y_{\tau})$, commonly referred to as ``climatology'' in meteorology (see definition in Box 1 and Figures \ref{fig:timeline}A and \ref{fig:timeline}B, Label 3) 
\citep{pappenberger_how_2015}. 
Since $\mathcal{R}$ represents the best available reference model capable of forecasting a system's future values, its scoring function, $\mathcal{S}( \varepsilon_t^\mathcal{R})$, must satisfy condition \ref{eq:ScoringInequation}, i.e.:
\begin{equation}\label{eq:condition_relative}
    \mathcal{S}\left( \varepsilon_t\right) \leq \varrho \leq \mathcal{S}\left( \varepsilon_t^\mathcal{R}\right) \ ,
\end{equation}
and as well as in section \ref{sec:formalisation}, when the last condition no longer holds ($\mathcal{S}(\varepsilon_t) = \varrho(t) = \mathcal{S}(\varepsilon_t)^\mathcal{R} $), the forecast limit is reached.
The ratio of such scoring functions, 
\begin{equation}\label{eq:relative_horizon}
\frac{\mathcal{S}(\varepsilon_t)}{\mathcal{S}(\varepsilon_t^{\mathcal{R}})} > 1\ ,
\end{equation}
defines a criterion to compare the score of the forecast model $\mathcal{M}$ to the reference model $\mathcal{R}$ \citep{pappenberger_how_2015}. 
Notably, if the left-hand side (LHS) of equation \ref{eq:relative_horizon} approaches 1, it indicates that the forecast produced by $\mathcal{M}$ performs as well as the one produced by the benchmark model $\mathcal{R}$. 
Conversely,  the more the LHS exceeds 1, the better the forecast performance of  $\mathcal{M}$. 
Hence, the first point in time where condition \ref{eq:condition_relative} no longer holds ($1 - \frac{\mathcal{S}(\varepsilon_t)}{\mathcal{S}(\varepsilon_t^{\mathcal{R}})} = \varrho =0 $) is the relative forecast limit, $h_r$ (see Figure \ref{fig:timeline} and \ref{fig:decision-tree}, Labels B), which is estimated as: 
\begin{equation}\label{eq:horizon_relative}
    h_r = \underset{t}{\text{argmin}} \left( 1 - \frac{\mathcal{S}(\varepsilon_t)}{\mathcal{S}(\varepsilon_t^{\mathcal{R}})}  \right) \quad \text{where}\quad t \in T_i \
\end{equation}

By definition, the expression in the brackets of \ref{eq:horizon_relative} is monotonic and
as shown in equation \ref{eq:step_function}, its evaluation can be expressed as another step function: 
\begin{equation*}
\label{eq:step_function_skill}
\gamma = 
\begin{cases} 
\text{1} & \text{if } \varepsilon_t > \varepsilon_t^{\mathcal{R}} \\
\text{0} & \text{if } \varepsilon_t < \varepsilon_t^{\mathcal{R}}
\end{cases}
\quad\text{where}\quad  t \in T_i 
\end{equation*}
which provides $h_r$ as $\gamma$ changes from 0 to 1. This approach is the \textit{forecast skill horizon} and defines the relative forecast limit (see Figure \ref{fig:timeline}B, Label B) \citep{buizza_forecast_2015}. Generally, the limit ${h}_r$ is only meaningful for comparisons when using the same verification $Y_t$ for scoring $\mathcal{M}$ and $\mathcal{R}$ with $\mathcal{S}$. 


\subsection{Potential predictability}\label{sec:potential_predictability}

When the realisable predictability is the lower bound of the model's forecast limit $h$, the model-intrinsic predictability is its upper bound. We introduce this as the potential forecast limit $\widehat{h}_p$, which determines the forecast limit of a theoretical system (see Figure \ref{fig:timeline} and \ref{fig:decision-tree}, Labels A) \citep{sun_intrinsic_2016, spring_predictability_2020}. 

The idea is based on the assumption that a variable is predictable as long as the climatological and the forecast distributions differ \citep{delsole_predictability_2007}. All components to equation \ref{eq:horizon_relative}, verification $Y_t$ and reference $\mathcal{R}$, are simulated from $\mathcal{M}$. Hence, $\mathcal{R} = \mathcal{M}$ generates a simulated climatological distribution, propagating uncertainties to any horizon, $T_i = t_i, ..., \tau$ where $t_i \ll \tau$.
This approach relies on sample ensemble forecasts, and their distribution is from the simulation rather than the observation as in equation \ref{eq:forecast_ensemble}. The forecast PDF then follows $p_{\widehat{Y}_{t,j}}({\widehat{Y}_t})$ where $j$ represents a randomly simulated trajectory from $\mathcal{M}$. Therefore, equation \ref{eq:probabilistic_error} becomes
\begin{equation*}
    \{ \varepsilon_{t}^{M} \} \sim p_{\widehat{Y}_{t,j}}({\widehat{Y}_t}).
    \label{eq:probabilistic_error_potential}
\end{equation*}

The potential forecast limit is then computed with the formalisation described in section \ref{sec:relative_skill}.

\subsection{Ad-hoc tolerance}\label{para:scoring_tolerance}

Using a reference model does not exclude the definition of a more rigorous $\varrho$. $\varrho$ may even be an ad-hoc expectation of forecast quality, e.g.~derived from literature \citep{petchey_ecological_2015}. In coupled Earth system models, the magnitude of predictive error on soil temperature that leads to unstable land-atmosphere interactions is known to be above 3 K \citep{zhou_evaluation_2024}. Another example is forecasts of forest productivity based on stand yield tables that classify forest productivity directly by the expected economic profit. In these examples, $\varrho$ has the unit of the forecasted state variable, e.g.~K for soil temperature or m for dominant stand height. We refer to a forecast limit that is computed with such an ad-hoc tolerance as the absolute forecast limit (see Figure \ref{fig:timeline} and \ref{fig:decision-tree}, Labels C).  When $\varrho$ is determined across groups within a grouping variable $k$ (such as species, see section \ref{sec:iLand} for an example), it becomes a function $\varrho(k)$ of that variable. 

\subsection{Scoring functions}\label{sec:scoring_functions}

We exemplify two scoring functions $\mathcal{S}$ that determine forecast limits from deterministic and probabilistic forecasts, following from equations \ref{eq:predictive_error} and \ref{eq:probabilistic_error}, respectively. 



\paragraph{Mean absolute error} The absolute error measures accuracy in the unit of the target variable without compensating positive and negative errors, and it is robust to outliers \citep{jolliffe_forecast_2003}. With the definition of the predictive error in equation \ref{eq:predictive_error}, it is at every time step defined as
\begin{equation}
    | \varepsilon_t |= |\widehat{Y_t} - Y_t\ | .
\end{equation}
When shifted by a fixed tolerance $\varrho$ that pre-defines acceptable accuracy ad-hoc, it behaves similarly to a scoring rule \citep{jolliffe_forecast_2003}, which is
\begin{equation}\label{eq:AE}
    \mathcal{S}_{\text{AE}_t} = \varrho - |\varepsilon_t |, \quad\text{where}\quad t \in T_i \,
\end{equation}
and the forecast limit is reached when $\text{AE}_t < 0$.
When $|\varepsilon|$ is computed for $m$ ensemble members or $m$ samples within a grouping variable and subsequently averaged, it collapses to the mean absolute error,
\begin{equation}\label{eq:MAE}
    \mathcal{S}_{\text{MAE}_t} = \frac{1}{m} \sum_{j=1}^{m} | \varepsilon_{t,i} |,  \quad\text{where}\quad t \in T_i \ \quad\text{and}\quad j \in \{1,\dots,m\}.
\end{equation}
For averaging of a distribution of predictive errors of ensemble members (see equation \ref{eq:probabilistic_error}), this is the mean ensemble error. This is not the error of the ensemble mean, which, following equation \ref{eq:probabilistic_error}, would be evaluated with \ref{eq:AE}.

\paragraph{Continuous ranked probability score}

The continuous ranked probability score (CRPS) evaluates a forecast probability density function (PDF), represented by the forecast ensemble, with regard to a scalar, representing the observed reference \citep[e.g.][]{hersbach_decomposition_2000, gneiting_comparing_2011}. When the ensemble PDF of the predicted target variable is $p(\widehat{Y})$ and the observed reference is $Y$, it is in discrete formalisation approximated within the finite interval $[\widehat{Y}_{min}, \widehat{Y}_{max}]$ as
\begin{equation}\label{eq:crps}
    \text{CRPS} = \text{CRPS}(\text{P}, Y) = \sum_{\widehat{Y}_{min}}^{\widehat{Y}_{max}} [\text{P}(\widehat{Y}) - H(\widehat{Y} - Y)]^2. 
\end{equation}
$\text{P}$ and $H$ are cumulative distribution functions (CDFs) that is the empirical CDF (often Gaussian) in the case of $\text{P}$ and that is the Heaviside function in the case of $H$. The Heaviside function for any value $x$ is a step function defined as 
\begin{equation*}
H(x) =
\begin{cases} 
\text{0} & \text{for } x < 0 \\
\text{1} & \text{for } x \geq 0 .
\end{cases}
\end{equation*}
$\text{P}$ is the associated CDF of the binary event that $\{Y \leq \widehat{Y}\}$ and is for any value $\widehat{Y}$ (in this context often termed ``threshold'') in discrete formalisation approximated as
\begin{equation*}
    \text{P}(\widehat{Y}) = \sum_{\widehat{Y}_{min}}^{\widehat{Y}} p(Y).
\end{equation*}

For a deterministic forecast, the CRPS generalises the MAE. As a stepwise skill score, it is evaluated as
\begin{equation}\label{eq:crpss}
    \mathcal{S}_{\text{CRPSS}_t} = 1 - \frac{\text{CRPS}_{t,\mathcal{M}}(\text{P}, Y_t)}{\text{CRPS}_{t,\mathcal{R}}(\text{P}, Y_t)}, \quad\text{where}\quad t \in T_i \,
\end{equation}
where it compares forecast and reference distribution \citep{buizza_forecast_2015}. It indicates perfect skill of $\mathcal{M}$ if $\text{CRPSS}_t = 1$ and skill of $\mathcal{M}$ no better than $\mathcal{R}$ if $\text{CRPSS}_t < 0$. 


\subsection{Monte Carlo uncertainty propagation}\label{sec:mc_sampling}

In two of the following three case studies, we use Monte Carlo simulation for propagating uncertainty in initial conditions, parameters, drivers, and/or the process itself into the forecast (Figure \ref{fig:decision-tree}, Label 4) \citep{dietze_ecological_2017}. The stochastic Ricker equation in case study 1 inherently represents uncertainty by stepwise simulating parameters from univariate Gaussian distributions with a standardised spread, pre-defined by the coefficient of variation (for details, see supplementary material, section 1). Initial condition uncertainty is propagated by a small perturbation of initial states, assuming a Gaussian error distribution. aiLand in case study 3 is a feed-forward neural network that has been trained with dropout layers. Dropout during inference can be interpreted probabilistically also for non-Bayesian neural networks and represents a measure of model uncertainty \citep{gal_dropout_2015}. When activated during inference, neurons in the network's hidden layers are randomly removed (set to 0) during forward simulation. In practice, the forecast distribution generated from these sources of uncertainties is received by repeated stochastic forward simulation.

\begin{tcolorbox}[breakable, title = Box 2. DIY forecast limits.]

To encourage their computation, we give a recipe on how to compute a forecast limit ($h$) within an experiment. To pick an approach that suits the experimental setting, the decision tree in Figure \ref{fig:decision-tree} suggests starting points for the choices we illustrated.

\paragraph{Step 1 (1): Select verification Y.} In hindcasting or near-term forecasting, these can be observations: Determine the relative forecast limit (Figure \ref{fig:decision-tree}, Label B). Without observational verification, $Y$ can be a simulated trajectory or the sample ensemble mean, generated from $\mathcal{M}$: Determine the potential forecast limit (Figure \ref{fig:decision-tree}, Label A). 

\paragraph{Step 2 (2): Select scoring reference $\varrho$.} Specify a scoring tolerance $\varrho$, a requirement for applying forecast limits. Best, use a reference model $\mathcal{R}$, i.e.~the climatological distribution. Rarely available, an ad-hoc tolerance $\varrho$ may be given based on a pre-defined expectation ~\citep[e.g.][]{petchey_ecological_2015, palamara_effects_2016}. 

\paragraph{Step 3 (3): Select scoring function(s) $\mathcal{S}$.} One or more functions $\mathcal{S}$ are required that evaluate the forecast relative to $\mathcal{R}$ or absolutely to the ad-hoc expectation (Figure \ref{fig:decision-tree}, Label 3). Depending on $\mathcal{S}$, the predictive error (equation \ref{eq:probabilistic_error}) will be evaluated as a distribution (CRPS) or point-wise (MAE).

\paragraph{Step 4 (4): Make forecast.} Run the forecast model to generate $\widehat{Y}^{\mathcal{M}}$. Use or create a reference model $\mathcal{R}$ to generate $\widehat{Y}^{\mathcal{R}}$. 

\paragraph{Step 4.1 (4): Propagate uncertainties.} In creating $\widehat{Y}^{\mathcal{M}}$ and $\widehat{Y}^{\mathcal{R}}$, propagate uncertainties from the components in $\mathcal{M}$, using e.g.~Monte Carlo sampling techniques or analytical error propagation (Figure \ref{fig:decision-tree}, Label 4) \citep{dietze_community_2023}.

\paragraph{Step 5: Step-wise evaluation.} Quantify forecast accuracy, precision, and/or skill iteratively over lead times with one or more scoring functions $\mathcal{S}$. Test if the score exceeds the tolerance $\varrho$ in step-wise evaluation. 

\paragraph{Step 6: Determine forecast limit.} Iteratively, decrease the forecast horizon $T_i$, or said differently, iterate over initial forecast times $t_i$. Repeat for multiple lead times $\tau$ and take the average. 

\begin{figure}[H]
    \centering
    \includegraphics[width=0.999\linewidth]{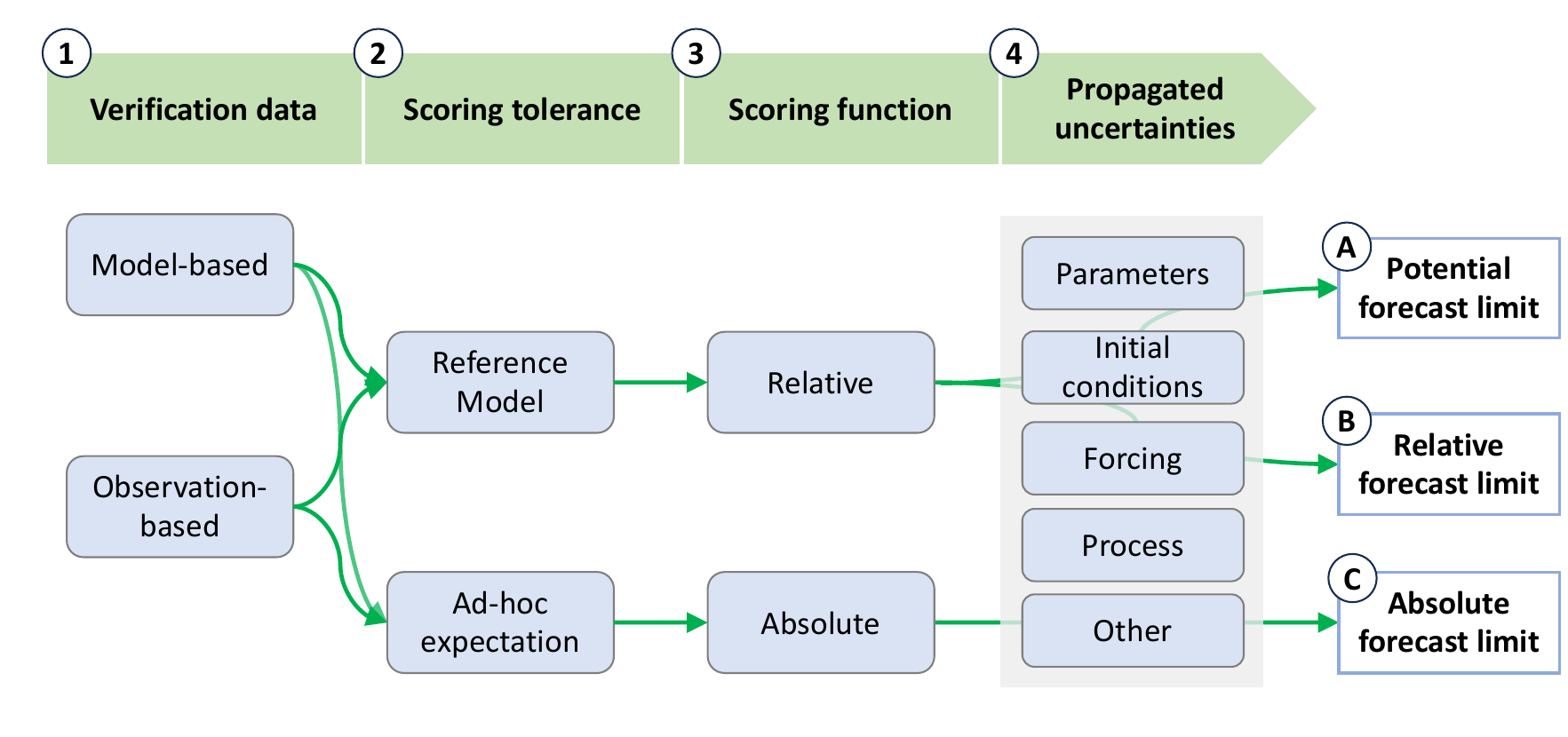}
    \caption{Decision tree for forecast limits: Depending on different choices of verification, scoring reference, and scoring function, this will result in one of the three classified types of forecast limit.}
    \label{fig:decision-tree}
\end{figure}

\end{tcolorbox}

\section{Results}

\begin{table}
    \centering
        \caption{Overview of the key characteristics of three different modelling approaches used in the case studies with the stochastic Ricker model, iLand, and aiLand, and of the choices of the computational components to compute their forecast limits.}
        \renewcommand{\arraystretch}{1.5}
    \begin{tabular}{p{1.7cm}l|p{3cm}|p{3cm}|p{3cm}}
        \multirow{1}{*}{} & & \textbf{Ricker} & \textbf{iLand} & \textbf{aiLand} \\ 
        \hline
        \multirow{3}{*}{\textbf{Model}} & \textit{Model type} & Mathematical & Process-based & Machine Learning \\ 
        & \textit{State variable} & Population size [-] &  Dominant height [m] & Soil temperature [K],\newline Soil moisture [$\frac{m^{3}}{m^{3}}$] \\ 
        & \textit{Driving dynamic} & Internal & External & External \\ \hline
        \multirow{4}{*}{\textbf{Forecast}} & \textit{Time scale} & Generational & Decadal & Seasonal-range \\ 
        & \textit{- representation} & Discrete & Discrete & Discrete \\ 
        & \textit{Spatial scale} & Local & Regional & Regional \\ 
        & \textit{- representation} & Single point & Multiple points & Multiple points \\ \hline
        \multirow{4}{*}{\textbf{Evaluation}} & \textit{Reference} & Simulations & Observations & Observations \\ 
        & \textit{Statistic} & Mean+Spread & Mean & Mean+Spread \\ 
        & \textit{Scoring function} & CRPSS & MAE & CRPSS\\ 
        & \textit{Tolerance} & Climatology & Ad-hoc & Climatology \\ \hline
        \multirow{1}{*}{\textbf{Horizon}} & & Potential & Absolute & Relative \\ 
    \end{tabular}
    \label{tab:model_types}
\end{table}

We demonstrate the computation of potential, absolute, and relative forecast limits in three case studies from population ecology, ecosystem ecology, and Earth system research. 
These case studies were selected to demonstrate the versatility of the forecast limit framework across different ecological contexts. The population ecological model represents a small-scale, internally driven, high-frequency dynamic, making it ideal for exploring the limits of predictability in stochastic and chaotic systems. The forest ecosystem model operates on decadal timescales and integrates complex external drivers such as climate and soil properties, providing an example of forecasting at larger spatial and temporal scales. Finally, the machine learning land surface model offers a cutting-edge application in forecasting, highlighting how the forecast limit concept can be applied to advanced, data-driven models. Our initial case study motivates the concept, while each subsequent one showcases how forecast limits can be determined and what experiment-specific choices are. The models vary in type, state variable, forecast range, and dynamic properties (see Table \ref{tab:model_types}), which is why we deliver the details on the model-specific experimental setup for determining forecast limits in this section rather than in the methods. However, in order to reduce information, we refer to the Appendix for detailed model descriptions and extended experiments.

\subsection{The stochastic Ricker equation and the potential forecast limit}\label{sec:ricker}

\begin{figure}[h]
    \centering
    \includegraphics[width=0.999\linewidth]{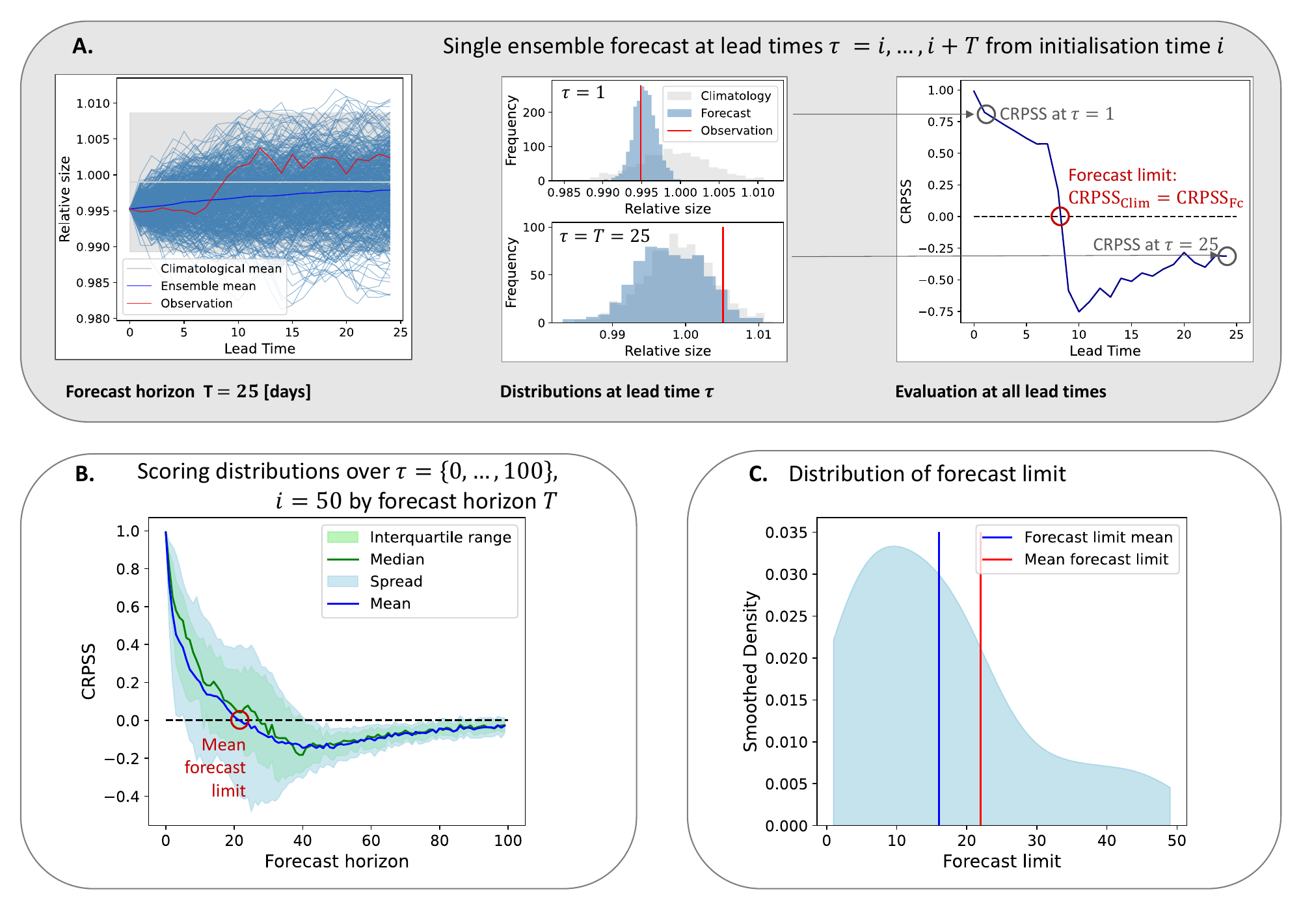}
    \caption{The forecast limit in a case study with the stochastic Ricker model. \textbf{A.} Ensemble forecast from one single initial forecast time $i=0$. Left: Generational lead time evolution of the steady state distribution of relative population sizes, propagating initial condition uncertainty and stochastic parameters uncertainty. The blue line indicates the ensemble mean, and the gray lines the climatological mean at the forecast horizon $T = 25$. The red line is the simulated observations. Middle: Forecast and climatological distribution at first lead time $\tau = 1$ and at horizon $\tau = T = 25$. Right: The CRPS of this forecast at all lead times. \textbf{B.} Distribution of the CRPSS by forecast horizon, representing an aggregation over lead times. The solid blue and green lines show the mean and median ensemble forecast limit, respectively. The light blue and green shade indicates the spread, i.e.~a single standard deviation of the ensemble forecast limit. \textbf{C.} The smoothed distribution of ensemble forecast limits. The red solid line shows the mean ensemble forecast limit as indicated in panel B, and the blue line the average of ensemble forecast limits.}
    \label{fig:ricker0}
\end{figure}

\paragraph{Experimental setup} 

The forecast model $\mathcal{M}(Y_0, X, \theta)$ is a single-species Ricker-type model that simulates a population dynamic. This is a discrete-time dynamical equation commonly used in theoretical community ecology to describe population growth \citep{ricker_stock_1954}. The estimated state variable $\widehat{Y}$ is the population size relative to an environment's carrying capacity, $k$, and the evolution of $\widehat{Y}$ over time represents the generational turnover of the population. We simulate a steady state dynamic at carrying capacity $k = 1$ in a non-chaotic parameter regime to reduce the complexity of the example.

With Monte Carlo error propagation, we iterate from $Y_0$ until $\widehat{Y}_\tau$ with a stochastic growth rate and carrying capacity with a $CV=0.03$ and initial conditions sampled from univariate Gaussian distributions with a $CV=0.001$. The stochastic parameters vary across time and are resampled at every time step. The observational verification $Y$ is simulated from that same model for demonstration without any process error, i.e.~the model perfectly describes the process. It can be understood as the one actually realised trajectory of the system in the real world. We mimic a forecast scenario where we predict from initial observations $Y_0$ over 25 generations ahead in the future in 1000 different simulations, to which individual runs we refer as \textit{ensemble members} (see Figure \ref{fig:ricker0}, A. left panel). The tolerance $\varrho$ is given by a benchmark model for which we use the model's ``climatological'' distribution (see Box 1 for explanation): This distribution is estimated from the saturated forecast distribution at verification time, forecasted from 1000 iterations \citep{delsole_predictability_2018}. It is assumed to be Gaussian with sample estimates from the saturated ensemble for mean and standard deviation.

Because we evaluate against a simulation as verification in a perfect model setting, we estimate the potential upper boundary of the forecast limit.
We then evaluate the model forecast repeatedly at all lead times, starting from different initialisation times $i = {0,..., 50}$. In doing so, we explore the relationship between skill and lead time and average out different verification times.

\paragraph{Results} 

First, we demonstrated how the forecast limit is determined in the example of a single ensemble forecast from initialisation time $i = 0$. As all simulations start at the same time, verification time is the same as lead time, i.e.~$\tau=25$ and $T=25$. Because at initiation time T=0 the system was below the carrying capacity of 1, the long-term trend is slightly upwards towards the climatological mean (= carrying capacity) for much longer runs (see Figure \ref{fig:ricker0}, A. left panel). The model predictive uncertainty, represented by the ensemble spread, increases over the forecast horizon, while the climatological distribution is constant. Initially, at $\tau_i \neq T_i$, the forecast distribution is sharp and centred around the initial observation until its mean and spread approach the climatological distribution at the maximum lead time $\tau = T = 25$ (see Figure \ref{fig:ricker0}, A. middle panel). The skill of the forecast distribution is evaluated towards the respective climatological distribution at all lead times, and the forecast limit is reached as the CRPSS drops below 0; this happens at generational verification time $\tau = 9$.

Second, we repeated the experiment over different initial forecast times, i.e.~verification times differ for the same lead times. From this experiment, we get two estimates for the model forecast limit (see Figure \ref{fig:ricker0}, B. and C.): One is the ensemble mean or median forecast limit, that is the average of the ensemble CRPSS over time (see Figure \ref{fig:ricker0}, B., blue and green solid lines). The mean forecast limit was estimated at 22 generations, with a 50\% confidence between 6 and 42 generations. Because the underlying distribution is not strictly normal but left-skewed, we also report the median forecast limit, which is at 28 generations, with upper and lower quartiles at 41 and 11, respectively. We receive a second estimate from the mean of ensemble trajectory forecast limits, which we termed the forecast limit mean (see Figure \ref{fig:ricker0}, C.). This was at 16.06$\pm$12.42 generations, yet again, the underlying distribution was non-normal, and the forecast limit median was at 13.5 generations with upper and lower quartiles at 20.75 and 6.25 generations, respectively.  

In summary, forecasts from this Ricker model reach their forecast limit at 9 time steps after initialisation based on CRPSS, while the forecast distribution allows for slightly longer forecasts.

\subsection{Forest growth with iLand and the absolute forecast limit}\label{sec:iLand}

\begin{figure}[h]
    \centering
    \includegraphics[width=0.999\linewidth]{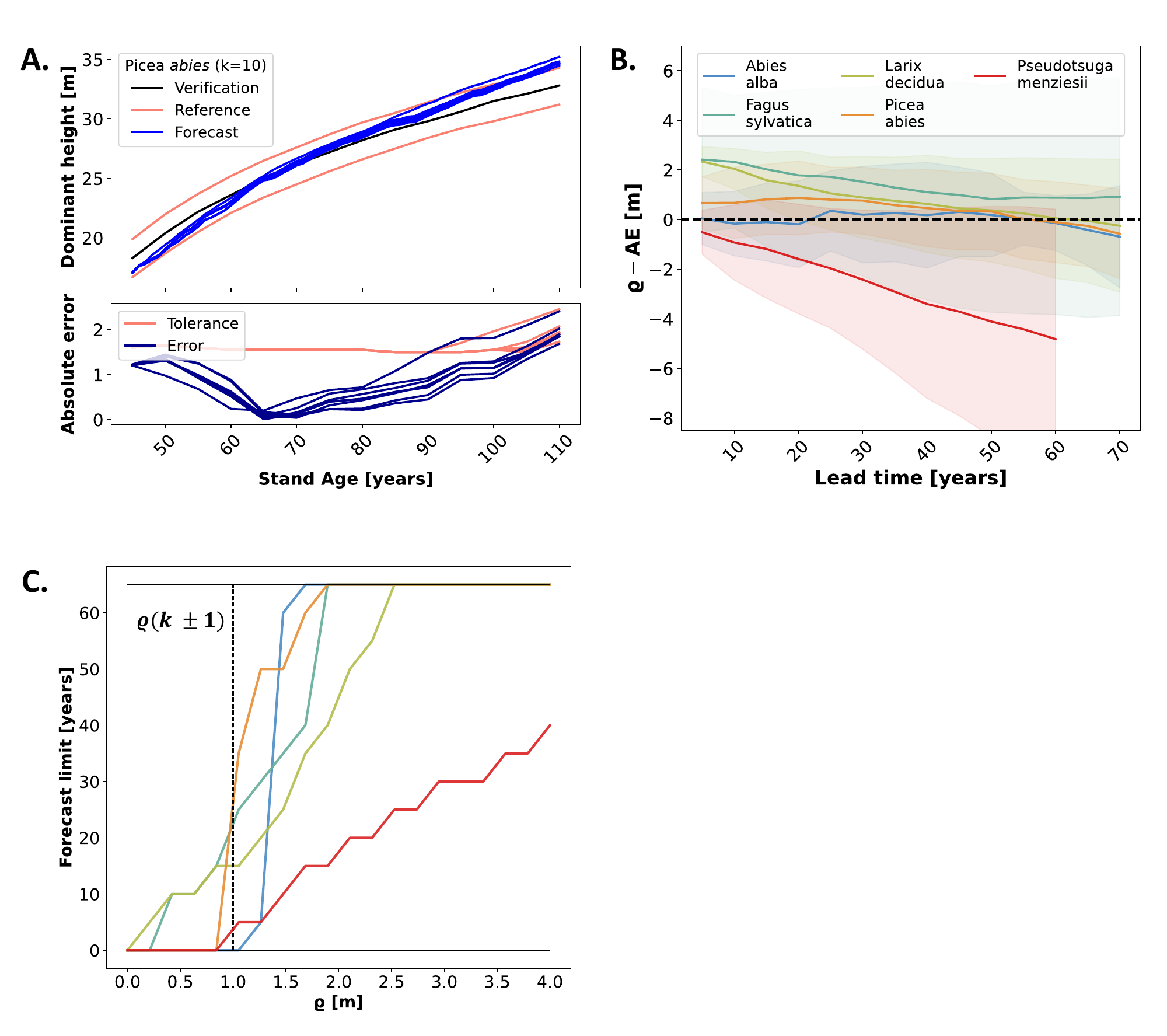}
    \caption{Absolute forecast limits for the dominant height of five tree species, forecasted with iLand, starting at a stand age of 45 years. \textbf{A.} Top: Exemplary growth trajectories of \textit{Picea abies} stands of observed yield classes $k = 10$. Reference growth trajectories (red) were taken from the regional yield tables (see definition of verification $Y$ in text). Bottom: Predictive error (dark blue) and tolerances (red) over forecasted stand age. \textbf{B.} The average forecast limit for all species. This is reached when the mean trajectories (solid lines) drop below 0. Mean trajectories were aggregated over stands and shaded areas, indicating the 95\% confidence interval caused by spatial variability. \textbf{C.} The forecast limit as a function of the tolerance $\varrho$ \citep{massoud_probing_2018}, independent of yield classes. The maximum possible limit is 65 years, referring to the maximum simulated stand age of 110.}
    \label{fig:iLand}
\end{figure}

\paragraph{Experimental setup} 

The forecast model $\mathcal{M}(Y_0, X, \theta)$ is the individual-based forest landscape and disturbance model iLand \citep{seidl_individual-based_2012, rammer_individual-based_2024}, which we use to demonstrate the absolute forecast limit of tree productivity. Primary tree productivity in iLand is represented, for example, by the stand dominant height (in meters), which we define as the system state $\widehat{Y}$, and which changes with stand age. The process is forced at daily resolution by four meteorological variables $X$ (temperature, precipitation, radiation, and vapour pressure deficit). Soil and carbon parameters are treated as global, constant parameters $\theta$. The study area was the Freiburger Stadtwald, which encompasses 269 sites and five tree species with varying numbers of observations. In this experiment, iLand made point-forecasts, i.e.~it produced a single forecast trajectory $\widehat{Y}_t$ for each site without error propagation. Each site trajectory was evaluated against the respective verification $Y_t$, and subsequently, the predictive errors were spatially aggregated across species. The verification $Y_t$ was an observation-inferred model of dominant stand height. For each stand, $Y_t$ was reconstructed from one-time forest inventory data, which was used to infer dominant height curves from regional yield tables for individual and species-specific \textit{yield classes}. Yield classes were identified and defined based on the productive capacity of a forest stand, specifically measuring its potential timber yield.  In Baden-Württemberg (region of case study), yield classification is determined through the average annual growth increment at 100 years of age ($dGZ_{100}$). The advantage of using this absolute yield class was that it standardised comparisons across tree species and yield classes by using the same reference age, improving comparability \citep[][p.~154f]{kramer_leitfaden_2008}. 

Using an ad-hoc expectation towards iLand's forecast performance, we estimated the absolute forecast limit (see Figure \ref{fig:timeline} and \ref{fig:decision-tree}, Labels C). This ad-hoc expectation was the reference model of each neighbouring yield class. Hence, $\varrho(t, k)$ was a function of yield class $k$ and of time $t$ and varied among stands and species. We chose a conservative reference of the next neighbouring yield classes, i.e.~if the observed yield class was $k$, the reference models were of yield classes $k\pm1$. The scoring function $\mathcal{S}$ was the absolute predictive error that collapses to the mean absolute error after spatial aggregation (see equation \ref{eq:MAE}). 

\paragraph{Results} 

iLand's absolute forecast limits varied notably both among and within species (see Figure \ref{fig:iLand}, Panel B). With our strict definition of $n=1$, average horizons were less than 75 years in stand age for \textit{Picea abies}, which corresponded to 35 years into the simulation; \textit{Pseudotsuga menziesii} even showed an average limit at stand age 45, corresponding to 0 years into the simulation. Yet, horizons varied strongly among stands and were long for \textit{Picea abies} and \textit{Fagus sylvatica}, where some stands show horizons until the maximum lead time. 
Showing the forecast limit as a function of the absolute error tolerance $\varrho$ (see Figure \ref{fig:iLand}, Panel C) is equivalent to shifting the dashed line in Panel B up or down. It reveals that a tolerance with two neighbouring yield classes, $\varrho(k)_{n=2}$, would extend the forecast limit of three species to the full lead time of 110 years.

\subsection{Land surface emulation and the relative forecast limit}

\begin{figure}[h]
    \centering
    
    \includegraphics[width=0.999\linewidth]{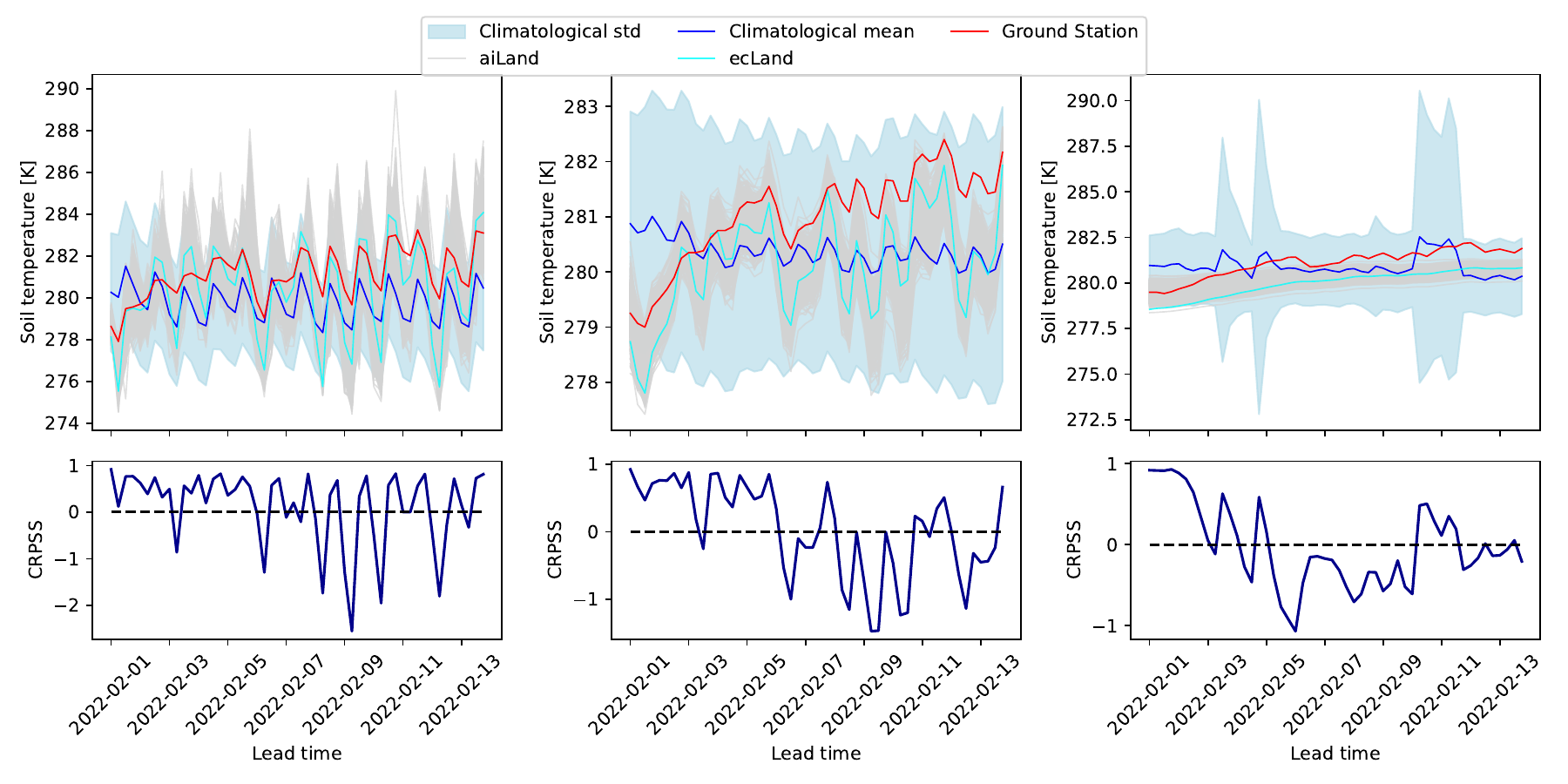}
    \begin{picture}(0,0)
        \put(-220,230){\textbf{A.}} 
    \end{picture}
    
    \includegraphics[width=0.999\linewidth]{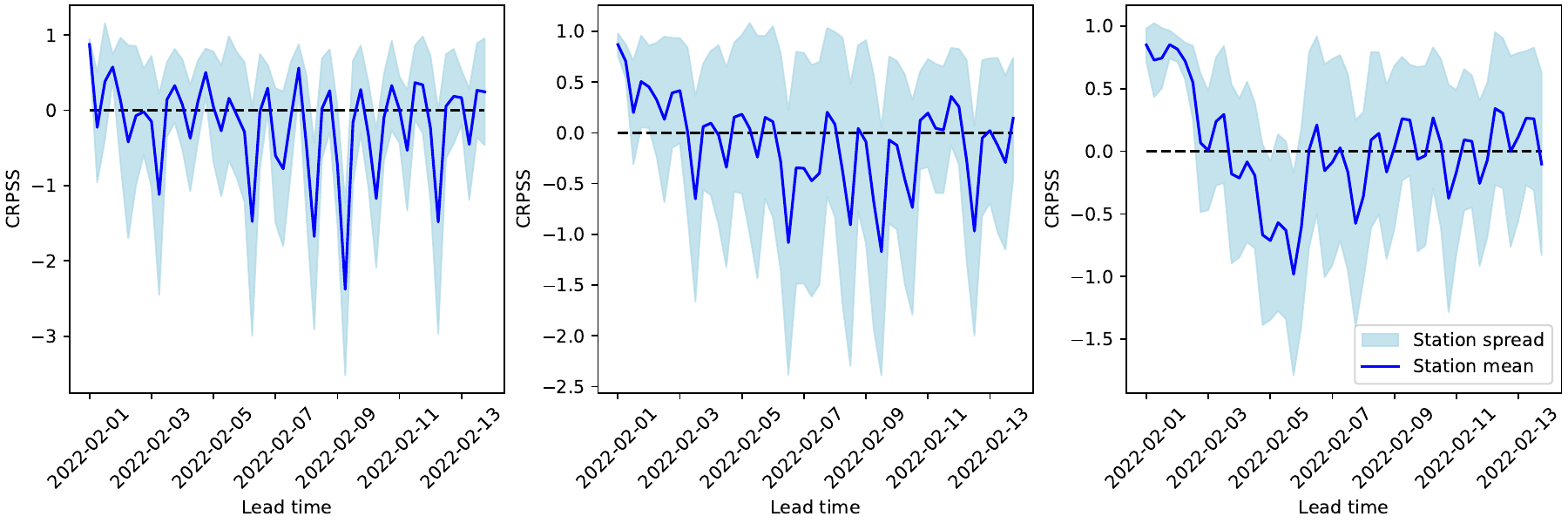}
    \begin{picture}(0,0)
        \put(-220,175){\textbf{B.}} 
    \end{picture}
    \caption{Soil temperature [K]. \textbf{(A)} Upper panel: Single, 6-hourly ensemble forecast over two weeks for one station (Condom-en-Armagnac) from 1) the aiLand (gray), propagating initial and model-structural uncertainty, and 2) the station climatological distribution (blue). aiLand was initialised with the ISMN station measurement on February 1st, 2022, 00:00:00. The non-initialised ecLand prediction is shown in turquoise. Lower panel: Model skill was determined relatively to the 6-hourly station climatological distribution with the continuous ranked probability score (CRPS). \textbf{(B)} The solid blue line indicates the station average (13 stations) of the CRPSS over all lead times and the light-blue shaded area the station standard deviation.}
    \label{fig:aiLand_temp}
\end{figure}

\begin{figure}[h]
    \centering
    
    \includegraphics[width=0.999\linewidth]{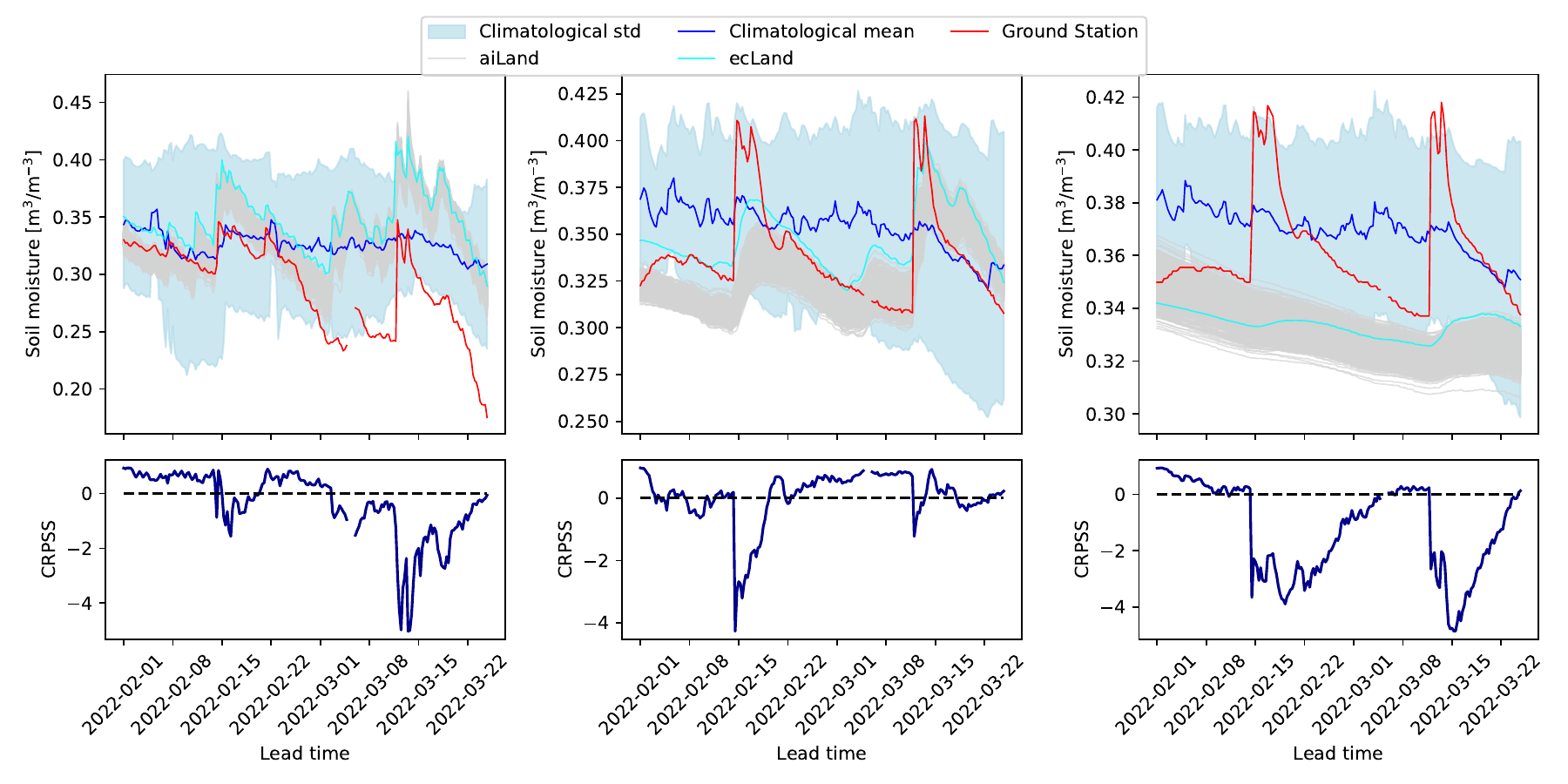}
    \begin{picture}(0,0)
        \put(-220,230){\textbf{A.}} 
    \end{picture}
    
    \includegraphics[width=0.999\linewidth]{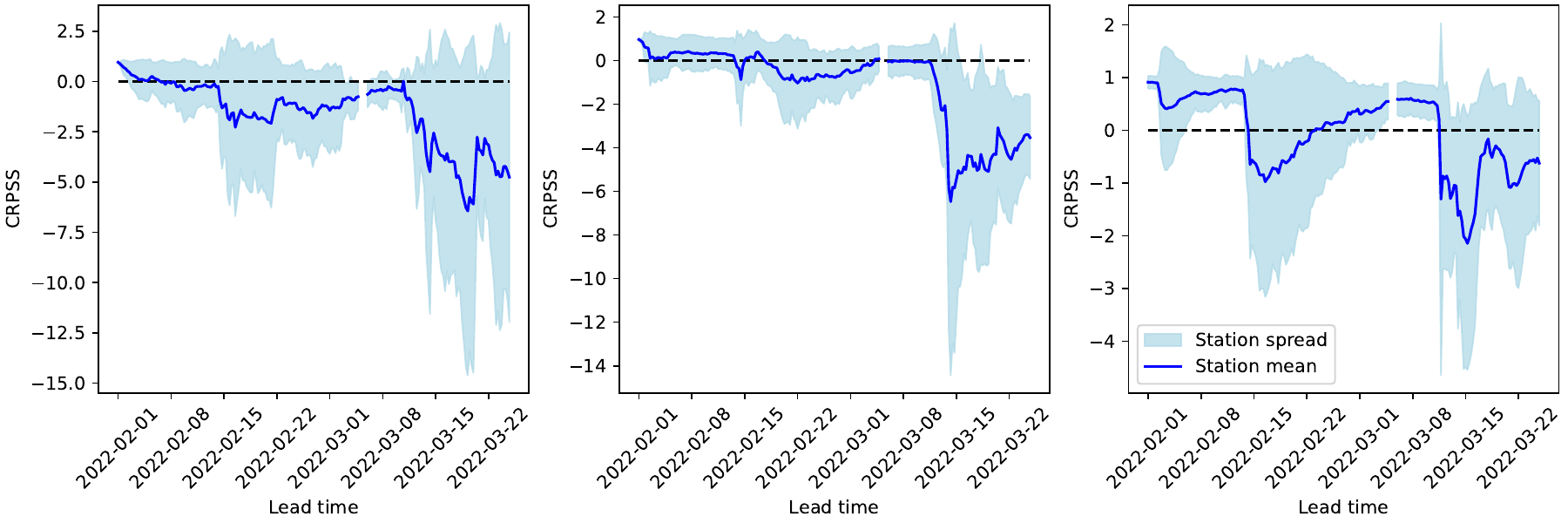}
    \begin{picture}(0,0)
        \put(-220,175){\textbf{B.}} 
    \end{picture}
    \caption{Soil moisture [m$^3$m$-^3$]. \textbf{(A)} Upper panel: Single, 6-hourly ensemble forecast over two weeks for one station (Condom-en-Armagnac) from 1), the aiLand (gray), propagating initial and model-structural uncertainty and 2) the station climatological distribution (blue). aiLand was initialised with the ISMN station measurement on February 1st, 2022, 00:00:00. The non-initialised ecLand prediction is shown in turquoise. Lower panel: Model skill was determined relatively to the 6-hourly station climatological distribution with the continuous ranked probability score (CRPS). \textbf{(B)} The solid blue line indicates the station average (7 stations) of the CRPSS over all lead times and the light-blue shaded area the station standard deviation.}
    
    \label{fig:aiLand_moist}
    
\end{figure}

\paragraph{Experimental setup} 

The forecast model $\mathcal{M}(Y_0, X, \theta)$ is a machine learning emulator of ECMWF's physical land surface scheme ecLand (hereafter: aiLand) \citep{boussetta_ecland_2021, wesselkamp_advances_2025}. Of the modelling approaches introduced in \citet{wesselkamp_advances_2025}, we here use only the multilayer perceptron to demonstrate the relative forecast limit \citep{buizza_forecast_2015}. State variables $\widehat{Y}$ of aiLand are soil water volume (i.e.~soil moisture, m$^3$m$^{-3}$) and soil temperature (K) at the soil surface layer (0-5 cm), subsurface layer 1 (5-20 cm) and subsurface layer 2 (20-70 cm), and snow cover. All states together initialise $\mathcal{M}$ as $Y_0$. 
External processes $X$ that force $Y$ are ERA-5 reanalysis dynamic meteorological variables and static climate and physiographic fields \citep{hersbach_era5_2020}. The details of the model and variables are described in \citet{wesselkamp_advances_2025}.
We show our analysis for soil temperature and soil moisture. aiLand was initialised and evaluated with measurements from the International Soil Moisture Network (ISMN) that provides a quality-controlled and harmonised database for land-surface process evaluation \citep{dorigo_international_2021}. We used local measurements from the years 2021 and 2022 of the French Smosmania network stations, which we defined as the reference $\textbf{Y}$. Soil temperature was measured at 0-5 cm (surface layer), 5-20 cm (subsurface layer 1), and 20-30 cm (subsurface layer 2). Due to missing measurements in January, aiLand was initialised with measurements on February 1st, 2022, constituting initial conditions $Y_0$.
Initial condition and in-sample model structural error were propagated with Monte Carlo sampling. First, initial states were perturbed in a naive approach, assuming an univariate Gaussian error distribution with a $CV=0.05$. Second, dropout layers were activated during model inference at a rate of 18\% (the training dropout rate, see \citet{wesselkamp_advances_2025}, supplementary material). Note that this version of aiLand had not been trained for probabilistic forecasting. A 1000-member ensemble was then integrated at a 6-hourly temporal resolution over $\tau = 1, \ldots, 52$ lead times for soil temperature, representing the medium-range, and over $\tau = 1, \ldots, 208$ lead times for soil moisture, representing the seasonal range. 

We computed the relative forecast limit of aiLand based on the measured station climatological distribution. This reference model $\mathcal{R}$ was assumed to be Gaussian and estimated from the long-term mean and standard deviation of all available 6-hourly measurements from 2008 to 2022. The scoring function $\mathcal{S}$ was the CRPS, evaluated against $\mathcal{R}$ as skill function CRPSS (see equation \ref{eq:crpss}). Equivalently to the case study in section \ref{sec:ricker}, we set $\varrho$ conservatively such that the limit was reached as performance of aiLand equals that of station climatology, i.e.~CRPSS$\leq0$. This was done for all stations.
For a preceding experiment with a deterministic evaluation and computation of an absolute and relative forecast limit of aiLand as a model ensemble, relatively to the physical model ecLand, we refer to the supplemental material.
As showcased in our case study in \ref{sec:ricker}, we repeated the analysis of forecast limits from different initialisation times for one station (Condom-en-Armagnac). This was done over a selected seasonal range of approximately nine weeks at 6-hourly resolution, starting again from February 1st, 2022. This reproduced the analysis that was conducted with the anomaly correlation coefficient in \citet{wesselkamp_advances_2025} relative to ecLand, but was here conducted with the CRPS relative to the ISMN measurements.

\paragraph{Results} 

\begin{figure}[h]
    \centering
    \includegraphics[width=0.95\textwidth]{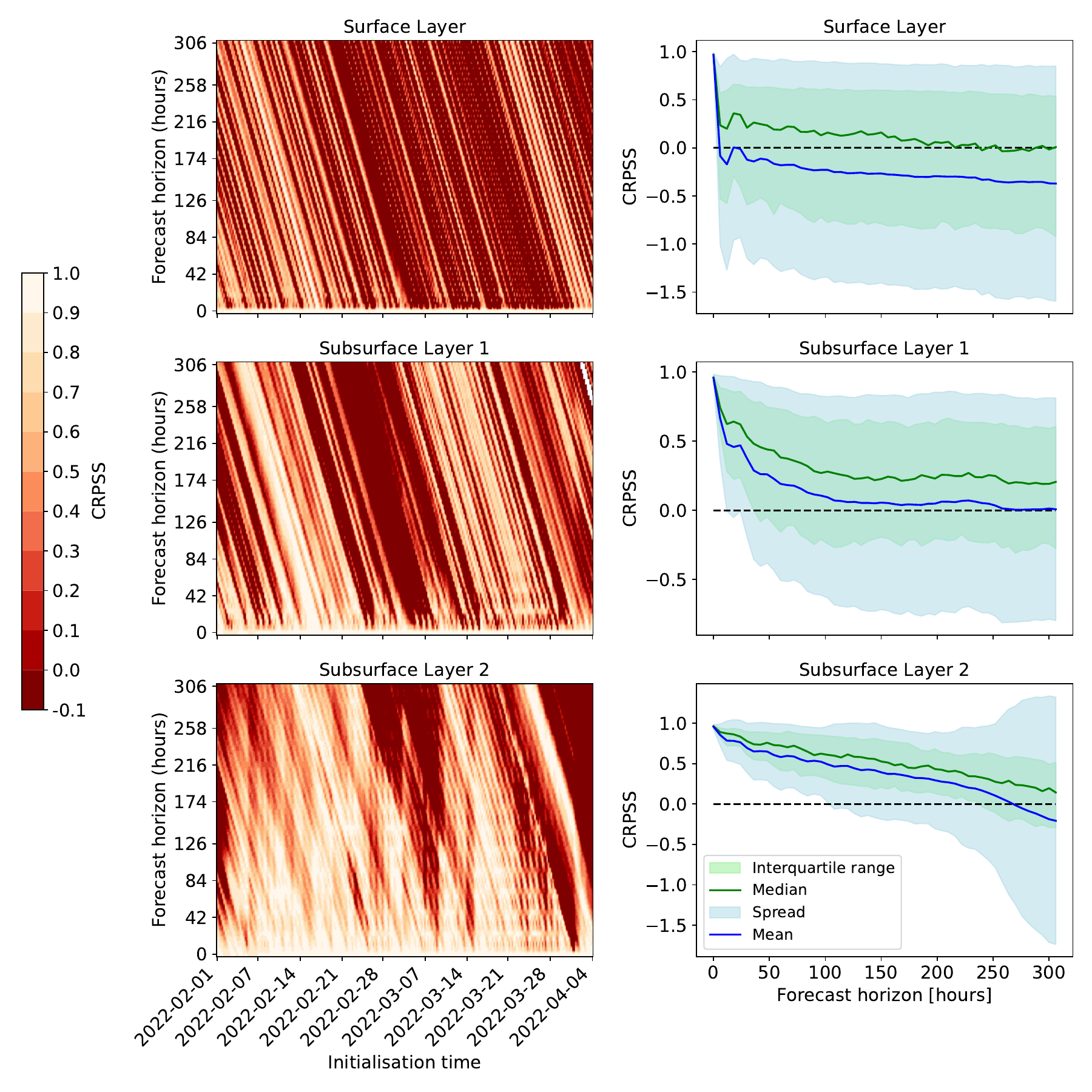}
    \begin{picture}(0,0)
        \put(-420,450){\textbf{A.}} 
        \put(-200,450){\textbf{B.}}
    \end{picture}
    \caption{Soil temperature [K]. \textbf{A.} Skill (CRPSS) over a forecast horizon of two weeks (312 hours) from varying initialisation times: Approx. nine weeks at 6-hourly resolution, starting on February 1st, 2022, 00:00:00. Light pattern indicates skilful forecasts, while dark pattern indicates no skill, relative to the climatological distribution. The oblique stripes, most evident for the surface layer, refer to the same forecast date, predicted at different lead times, i.e. from different initialisation times (analysis reproduced from \citet{wesselkamp_advances_2025}). \textbf{B.} Lead time mean (blue) and median (green) of the CRPSS across initialisation times for all forecast horizons. This corresponds to a bottom-to-top, row-wise average of the skill displayed in  \textbf{A} (see also Figure \ref{fig:ricker0} B). The spread (i.e. a single standard deviation) is indicated by the light-coloured shadows.}
    \label{fig:aiLand_horizons_temp}
\end{figure}

\begin{figure}[h]
    \centering
    \includegraphics[width=0.95\textwidth]{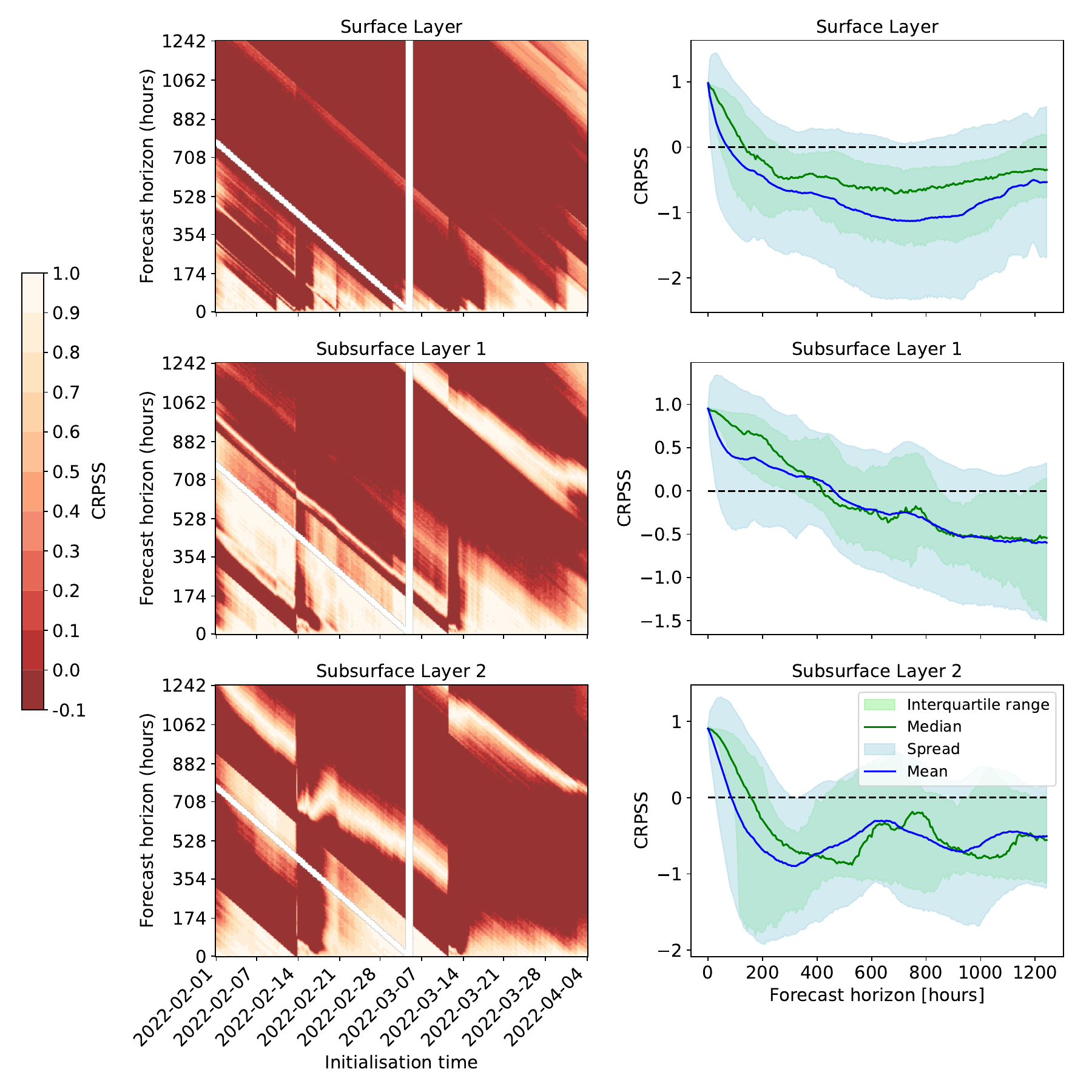}
    \begin{picture}(0,0)
        \put(-420,450){\textbf{A.}} 
        \put(-200,450){\textbf{B.}}
    \end{picture}
    \caption{Soil moisture [m$^3$m$^{-3}$]. \textbf{A.} Skill (CRPSS) over a forecast horizon of eight weeks (1248 hours) from varying initialisation times: Approx. nine weeks at 6-hourly resolution, starting on February 1st, 2022, 00:00:00. Light pattern indicates skilful forecasts, while dark pattern indicates no skill, relative to the climatological distribution. The oblique stripes refer to the same forecast date, predicted at different lead times, i.e.~from different initialisation times (analysis reproduced from \citet{wesselkamp_advances_2025}). The white stripes indicate missing measurements. \textbf{B.} Lead time mean (blue) and median (green) of the CRPSS across initialisation times for all forecast horizons. This corresponds to a bottom-to-top, row-wise average of the skill displayed in \textbf{A} (see also Figure \ref{fig:ricker0} B).}
    \label{fig:aiLand_horizons_moist}
\end{figure}

Using the example of a single station, it is evident that the aiLand ensemble, which accounts for both initial and structural errors, exhibits relatively small and consistent variability over the forecast horizon for both soil temperature and moisture (see Figure \ref{fig:aiLand_temp} A and \ref{fig:aiLand_moist} A), with a tendency to converge at long lead times. The measured climatological average was for both variables relatively steady with a large spread, compared to the 2022 measurements, specifically on the surface layer and first subsurface layer. 
Initialising aiLand with station measurements resulted in a short forecast limit for soil temperature, determined with the CRPSS relative to the climatology. Averaging over all stations, it was especially short on the surface layer, where the CPRPSS varies at high frequencies around the threshold of 0 with a slight negative tendency over the explored time period (see Figure \ref{fig:aiLand_temp} B.). The forecast limit extended towards deeper layers (this resembled results from the deterministic analysis relative to ecLand, see supplementary material). At all depths, the forecast limit was longer for soil moisture than for soil temperature (see Figure \ref{fig:aiLand_moist} B). Yet, for the station example in Figure \ref{fig:aiLand_moist} A, the modelled dynamic showed qualitative mismatches with the local measurements at the second subsurface layer. 

Based on the analysis from \citet{wesselkamp_advances_2025}, the stepwise evaluation was iteratively repeated from different initialisation times over approximately nine weeks for a single station (Condom-en-Armagnac). The resulting heat maps in Figures \ref{fig:aiLand_horizons_temp} and \ref{fig:aiLand_horizons_temp} A reveal whether the relative predictability is dependent on the lead time or dominated by seasonality. The oblique strips indicate the CRPSS of aiLand relative to the climatological distribution at the exact dates, forecasted from different leads, i.e.~at different lead times. Their inclination only tells us about the number of forecast horizons and different initialisation times. Colour gradients along the oblique stripes indicate a dependency of model quality on lead time and result in light patches and triangles (see Figures \ref{fig:aiLand_horizons_temp} and \ref{fig:aiLand_horizons_temp} A, subsurface layers).

The relative forecast limit was re-estimated after averaging over lead times, although such an approach is less meaningful when external drivers or seasonality dominate the dynamic, which often is the case for land surface processes. This was evident for soil temperature in Figure~\ref{fig:aiLand_horizons_temp} A, surface layer, where a strong day-to-day variability obscured any dependency on lead time, which resulted in a short forecast limit of, on average, just 6-hours, i.e.~one time step (see Figure \ref{fig:aiLand_horizons_temp} B, surface layer). For the surface layer's soil temperature, the forecast horizon was a poor predictor of model skill. This was different for temperature in the lower soil layers, where the lead time average in skill declines more slowly. The model did not reach the forecast limit in the first subsurface layer and approached it after an average of 11 days on the second subsurface layer (see Figure \ref{fig:aiLand_horizons_temp} B). 
The relative forecast limit for soil moisture was more dependent on lead time and hence generally better determined by the forecast horizon (see Figure \ref{fig:aiLand_horizons_moist} A and B). The results indicated a mean forecast limit for soil moisture on the surface layer of three days, of approximately 20 days on the first and approximately four days on the second subsurface layer (Figure \ref{fig:aiLand_horizons_moist} B), with strong variability.


\section{Discussion}

Our intention was to highlight the relevance of quantified forecast limits and improve credibility when communicating model-based forecasts. We motivated, distinguished and demonstrated three ways of determining empirical forecast limits that we termed potential, absolute, and relative forecast limits. 

\subsection{Motivating forecast limits}

Mechanistic statements about random processes will always include uncertainty, and ideally, forecasts make statements about the future with full uncertainty propagation \citep{clark_ecological_2001, dietze_ecological_2017}. The forecast limit is a concept that falls into the realm of forecast verification and has been introduced before, purely theoretically \citep{petchey_ecological_2015}. As defined in section \ref{sec:formalisation} and illustrated in section \ref{sec:ricker}, it evaluates forecast skill as a function of lead time, relative to a scoring tolerance ($\varrho$) for model quality. This tolerance is assumed to be a reference model, or, in rare cases, an ad-hoc expectation towards predictive error. The forecast horizon refers to the time period over which we forecast, and it is ignorant of the uncertainties that obscure the model statement. The forecast limit, on the other hand, accounts for the uncertainty that affects the system's predictability and for forecast quality decreasing with increasing uncertainty.
The predictability of any stochastic Markov system declines over time \citep{guo_rebound_2012}, when predictability is defined based on the dispersion of the forecast distribution, i.e.~the noise or spread \citep{delsole_predictability_2018}. This statement assumes that a larger spread of the forecast distribution leads to a larger predictive error, i.e. positive spread-error correlation, but this depends on how spread and error are quantified \citep{hopson_assessing_2014}. Ideally, the demonstration of forecast limits will be extended in future studies with full error propagation for each case study and the respective spread-error correlation being assessed.


\subsection{Potential, absolute and relative forecast limits}

\paragraph{Stochastic Ricker} 

In this simulation study, we demonstrated the forecast limit as the potential limit of predictability of a Ricker system in a steady state, propagating initial conditions and parameter uncertainty. While the evaluation procedure is equivalent to that of the relative forecast limit, the \textit{potential} forecast limit estimates the upper boundary of predictability by using a model simulation as a verification \citep{sun_intrinsic_2016, spring_predictability_2020}.  
This highlights how a limit of system predictability is determined from forecast uncertainty. In the potential forecast limit, this happens relative to the model-intrinsic uncertainty that is represented by the saturated ensemble forecast distribution. The forecast limit is reached as the ``climatological'' (long-term) distribution and the forecast distribution are indistinguishable. We expect the forecast limit to be sharper when additional sources of uncertainty are present, such as observation or process errors.

\paragraph{iLand} In the iLand case study, we made deterministic point forecasts without uncertainty propagation, only considering the impact of forcing variability on the absolute forecast limit through spatial aggregation. Stand productivity in iLand is a stable state variable whose absolute forecast limit is determined by external standards. While this approach implicitly touches on forecast utility through yield classes that represent economic value \citep{murphy_what_1993}, we evaluate forecast quality. iLand is commonly used for projections, but the framework we employ requires reference data to determine the limit of a point forecast. Ideally, observational data would serve as the primary reference. However, with only three inventory measurements per stand over a 30-year span, we relied on reconstructed observations from yield tables as the most viable alternative, which, strictly speaking, are also model-based (see description in the Appendix, Case Study 2); thus, this analysis cannot be used to draw definitive conclusions about iLand, but it demonstrates what we defined as \textit{absolute} forecast limit (\cite[as introduced by][]{petchey_ecological_2015}). This definition relies on specifying a tolerance towards the predictive error based on pre-defined standards.

\paragraph{aiLand} 


In the aiLand case study, a feed-forward neural network emulated the ECMWF's land surface scheme ecLand \citep{boussetta_ecland_2021, wesselkamp_advances_2025}. We propagated model structural uncertainty with Monte Carlo dropout \citep{gal_dropout_2015} and initial conditions uncertainties by a naive perturbation and determined a forecast limit relative to the climatological measurement distribution with the CRPSS \citep[e.g.][]{hersbach_decomposition_2000, gneiting_comparing_2011}. However, aiLand was not trained probabilistically but with a first-order loss function that optimises towards the mean. Deterministically trained neural networks have been shown to underrepresent variability in meteorological forecasts \citep{bonavita_limitations_2024}. On the other hand, being externally driven, we expect initial conditions uncertainty to converge over time \citep{dietze_iterative_2018}.

Our results indicated a short forecast limit of soil temperature of the surface layer. This limit, however, was not only dependent on the forecast model, but also on the reference model and the suitability of the forecast task. The reference model for temperature and moisture was the 6-hourly distribution of twelve years of station measurements, approximated with a normal distribution. Ideally, multiple reference models and multiple metrics would be compared \citep{pappenberger_how_2015}; for example, persistence may be a suitable reference model for soil moisture. Furthermore, the forecast is not just limited by model structural errors, but also by uncertainty in meteorological forcing. aiLand was trained on and forecast to a 31 km spatial grid, while the verification is based on station measurements, and local conditions may dominate temperatures of the surface layer. As such, if aiLand was driven by erroneous forcing, this will likely decrease the forecast limit.

\subsection{Forecast limits beyond chaotic systems}

The forecast limit was initially defined for sub-seasonal weather forecasting of the endogenously unstable system of the atmosphere \citep{lorenz_predictability_1996}.  
The instability of a system puts an upper bound on its predictability and is caused by sensitivity to initial conditions, which can lead to the exponential growth of predictive errors \citep{lorenz_atmospheric_1982, dalcher_error_1987}. Unstable dynamics \citep{rogers_chaos_2021} or alternative stable states \citep{scheffer_catastrophic_2001} may exist in natural ecosystems, but, as open systems, they are externally driven by meteorological forcing \citep{jorgensen_ecosystem_2009}. This is also well known for hydrological forecasts, which, driven by medium-range weather forcing, have an estimated predictability limit of ten to 14 days \citep{zhang_what_2019, bogner_tercile_2022}. 

However, recent insights suggest sensitivity to initial conditions in decadal forecasting of biomass in forest ecosystems \citep{raiho_towards_2020}. Here, the sensitivity is caused by the long memory of vegetation and soil, processes related to carbon and nitrogen pools that initialise a model and which are at best constrained by observations. This was also the case for iLand, and in the current study we did not explore how initial conditions uncertainty affects the forecast limit of iLand. However, a full predictability analysis following \citet{raiho_towards_2020} could greatly support an enhanced understanding of predictable timescales in landscape-scale forest dynamics.
The land surface processes represented in aiLand are forced processes, strongly driven by meteorological conditions. Already on the medium to sub-seasonal ranges (see Figures \ref{fig:aiLand_temp} and \ref{fig:aiLand_moist}), we observe the convergence of initial state perturbations. The forecast limit on the land surface thus depends on the predictability of and sensitivity to forcing; it is limited by seasonality and therefore is strongly location-dependent \citep{doblasreyes_seasonal_2013, bogner_tercile_2022}. This characteristic makes lead time averages over seasons less meaningful, or at least less interpretable than for internally driven systems. In seasonal forecasting of forced processes, the forecast limit can be reached temporarily. Areas of low predictability can arise during seasonal changes when forcing variability dominates and forecast performance is independent of lead time \citep[e.g.][]{wesselkamp_advances_2025}. However, the model can regain skill at longer lead times, a phenomenon that has been called ``the return of skill'' \citep{guo_rebound_2012}.

\subsection{Contributions to ecological forecasting}

We propose establishing forecast limits to monitor ecological forecasting systems and enhance the tractability of model-based statements in ecology for both the scientific community and decision-makers \citep{clark_ecological_2001}. Forecast limits enable tracking the skill development of forecasting systems \citep{bauer_quiet_2015}, and they depend on the predictive uncertainty, caused by multiple sources \citep{dietze_iterative_2018}. Therefore, they are not fixed but can be improved upon, depending on the uncertainty that drives them. Forecast limits establish quantitative assessments of forecast skill as a function of lead time, and establishing them will support systematic benchmarking and identification of paradigm shifts or significant advancements in ecological forecasting. 

Based on the previous state of discussion on forecast limits \citep{buizza_forecast_2015, petchey_ecological_2015}, we focused on three requirements for their computation: a verification, be it observation-based, model-based, or mixtures thereof, an appropriate scoring function, and a reference for evaluation that defines a predictive error tolerance. We extended previous works through the formal framework and differentiated three approaches to studying forecast limits that we term potential, absolute, and relative estimates of predictability \citep{buchovecky_potential_2023, tiedje_potential_2012, sun_intrinsic_2016, buizza_forecast_2015}: 
The \textit{relative} forecast limit assesses the realised predictability of the forecast relative to a benchmark or null model \citep{buizza_forecast_2015}. It is useful for model comparisons of any type when a reference model is reported \citep[e.g.~as in][]{wheeler_predicting_2024}, and we demonstrated it in the aiLand case study.
The \textit{potential} forecast limit assesses the internal predictability of the modelled system and can be interpreted as the upper boundary estimate, conditional on the model. It is applicable without an observational verification or the need to specify an ad-hoc predictive error tolerance. We refer to the \textit{absolute} forecast limit those approaches that, instead of using a best available reference model, use a pre-defined standard towards error tolerance, motivated by system-specific requirements \citep{petchey_ecological_2015}. This can depend on time and a variable that represents the ad-hoc standard.

\subsection{Directions from here}\label{sec:directions}

When forecast limits are established to quantify predictability and guide relevant time-scales for ecological forecasting and decision-making, there will be a space for studying the requirements to compute them. In this work, we used two metrics for evaluation (MAE and CRPS) and one reference model (lead-time climatology or long-term mean) across case studies, and we distinguish among types of verification. Optimally, multiple scores and reference models would be compared within a single study. This would help determine appropriate reference models, which depend on the forecasted variable and thereby improve model comparisons \citep{pappenberger_how_2015}. 
Further, in ensemble forecasting, it is important to consider the spread-error relationship that measures how well the dispersion of the forecast ensemble reflects forecast uncertainty. This relationship is defined as the correlation between predictive error of the ensemble mean and the ensemble spread, and it is assessed with coherent statistics, e.g., the mean squared error of ensemble mean and verification, and the standard deviation of ensemble members, respectively \citep{hopson_assessing_2014}. Their relationship is used to assess the reliability of predictions: A strong correlation between spread and error indicates that the model ensemble can reliably estimate forecast uncertainty. Incorporating this relationship deeper into the evaluation of forecast limits could provide a more robust assessment of model reliability.
Finally, as forecast limits depend on the growth of predictive uncertainty over the forecast time, partitioning the sources that contribute to this uncertainty will indicate how the forecast limit can be extended \citep{raiho_towards_2020, dietze_near-term_2024}. This could be studied theoretically in experiments that explore the forecast limit as a function of contributions to uncertainty across different sources.



\subsection{Limitations}

We explored the computation of forecast limits for their requirements across case studies. In two of the case studies, we propagate selected uncertainties, while in one case study we evaluate a point forecast. However, the forecast limit is influenced by all sources of uncertainty that affect model and forecast accuracy as well as the sensitivity to each source \citep{dietze_iterative_2018}. This includes driver uncertainty, model fit, parameter variability, ensemble size, and initial condition uncertainty and sensitivity. A full predictability assessment would involve propagation of all these sources of uncertainty into $\varepsilon$ \citep{dietze_prediction_2017, raiho_towards_2020}. 

Furthermore, the simplified distinction we take on model- and observation-based verifications is theoretical and neglects the error on observations themselves, arising through the unknown latent ``true'' state of the system. Further, when using observations in real-world scenarios, most of the time models are involved already and available validation data are mixtures \citep{edwards_vast_2013}. Their differentiation and the typology of forecast limits need to be clarified in future studies.

We want to stress that forecast limits are conditional on the forecasting model $\mathcal{M}$ (see \ref{eq:forecast_model}) and thus they do not represent the ``true'' limits of predictability. The potential forecast limit may be a good estimate of the relative forecast limit, but this is system-specific \citep{delsole_statistical_2022}, depending on how well $\mathcal{M}$ captures the features of the system. Moreover, if a different model were used to forecast the same process, we would likely find a different estimate for the limit \citep{shen_lorenzs_2023}. Model ensembles, such as in our experiment with aiLand in the supplementary material, can be used to account for this type of uncertainty. 

Finally, the methods we applied and the terminology we introduce in large parts descend from studies on operational forecasting systems. While we conducted theoretical forecasting studies, technically in hindcasting settings, relating the discussed concepts closer to forecasting studies will further clarify their application. However, our approach to determine forecast limits is based on forecast quality, not on forecast utility to the user \citep{murphy_what_1993, jolliffe_forecast_2003}. Determining forecast limits based on utility would be a non-trivial undertaking, as utility is linked to the value of a forecast to its user for decisions on action-taking \citep[][p.~165]{jolliffe_forecast_2003}, and hence not directly accessible to the forecaster. 

\section{Conclusion}

The forecast limit can be defined and applied to practically any ecological forecast. It facilitates interpretation and communication of forecast quality by evaluating skill as a function of lead time. We demonstrated the concept theoretically and applied it in three case studies. Our case studies represent scenarios with models that are used in operational forecasting (aiLand), models that are used for decision making (iLand) in a field where predictive ecology is already practised successfully \citep{evans_predictive_2013}, and models that are used to explore hypotheses qualitatively and theoretically (Ricker) \citep[e.g.][]{palamara_effects_2016}.
In our framework, determining forecast limits with scoring functions requires a verification, which can be observations or model simulations, and a reference for the score, which can be a null model or an ad-hoc value. Choices of the verification and the reference for scoring tolerance lead to different types of forecast limits (absolute, relative, potential). We suggest that ecological and, indeed, environmental forecasts should be accompanied by computations of forecast limits relative to reference models when observations are available as verification and especially when multiple forecast models are compared. Reference models and lead time evaluation are already used as benchmarks for comparative evaluation of ecological forecasts submitted to the NEON (National Ecological Observatory Network) forecasting challenge \citep[e.g.][for plant phenology forecasts]{wheeler_predicting_2024}, providing a pre-requisite for reporting forecast limits. Furthermore, we suggest reporting the potential forecast limit in model-specific studies to explore the system's behaviour under uncertainty. 
Future studies will have to evaluate the appropriate time scales, reference models, and scoring functions for specific ecological forecasting tasks; for now, following our suggested approaches can be seen as a piece of complementary information to the forecast itself. 

\section*{Code and Data availability}

\textbf{Code and Data availability}: ISMN data is available via \url{https://ismn.earth/en/} \citep{dorigo_international_2021}. The aiLand database is available via \url{https://doi.org/10.21957/pcf3-ah06} \citep{wesselkamp_advances_2025}. Forcing data and parameters for iLand, the archived code and remaining necessary data are available via \url{https://osf.io/rtu2f/?view_only=24e1241fc62140ca93589138bbaa934d} \citep{razafimaharo_new_2020, poggio_soilgrids_2021}.

\section*{Author Contributions}
Marieke Wesselkamp, Florian Pappenberger and Carsten F.~Dormann conceived of the study. Marieke Wesselkamp and Jakob Albrecht conducted the analysis. Marieke Wesselkamp and William J.~Castillo wrote the formalisation. EP provided comments and evaluation data bases. Florian Pappenberger, Carsten F.~Dormann and William J.~Castillo reviewed the study. Marieke Wesselkamp, Jakob Albrecht, Florian Pappenberger, William J.~Castillo and Carsten F.~Dormann wrote the manuscript.

\section*{Acknowledgments}
This work profited from helpful discussions with Souhail Boussetta, Christoph Rüdiger, Gabriel Moldovan and greatly from the comments of two anonymous reviewers. We gratefully thank Margarita Choulga and David Fairbairn for providing pre-processed ISMN data, and the FVA-BW for the BW-Inventory Data. We thank ECMWF for computational resources, ERA5 data and climate fields, and financial support for the research visits of MW. CFD acknowledges partial funding by DFG CRC 1537 Ecosense. 

\section*{Conflict of interest}

The authors declare no conflict of interest.

\bibliographystyle{apalike} 
\bibliography{forecast-horizons}

\begin{thebibliography}{}

\bibitem[Adler et~al., 2020]{adler_matching_2020}
Adler, P.~B., White, E.~P., and Cortez, M.~H. (2020).
\newblock Matching the forecast horizon with the relevant spatial and temporal processes and data sources.
\newblock {\em Ecography}, 43(11):1729--1739.

\bibitem[Albrich et~al., 2018]{albrich_tradeoffs_2018}
Albrich, K., Rammer, W., Thom, D., and Seidl, R. (2018).
\newblock Trade‐offs between temporal stability and level of forest ecosystem services provisioning under climate change.
\newblock {\em Ecological Applications}, 28(7):1884--1896.

\bibitem[Bauer et~al., 2023]{bauer_deep_2023}
Bauer, P., Dueben, P., Chantry, M., Doblas-Reyes, F., Hoefler, T., McGovern, A., and Stevens, B. (2023).
\newblock Deep learning and a changing economy in weather and climate prediction.
\newblock {\em Nature Reviews Earth \& Environment}, 4(8):507--509.

\bibitem[Bauer et~al., 2015]{bauer_quiet_2015}
Bauer, P., Thorpe, A., and Brunet, G. (2015).
\newblock The quiet revolution of numerical weather prediction.
\newblock {\em Nature}, 525(7567):47--55.

\bibitem[Ben~Bouallègue et~al., 2024]{ben_bouallegue_rise_2024}
Ben~Bouallègue, Z., Clare, M. C.~A., Magnusson, L., Gascón, E., Maier-Gerber, M., Janoušek, M., Rodwell, M., Pinault, F., Dramsch, J.~S., Lang, S. T.~K., Raoult, B., Rabier, F., Chevallier, M., Sandu, I., Dueben, P., Chantry, M., and Pappenberger, F. (2024).
\newblock The rise of data-driven weather forecasting: {A} first statistical assessment of machine learning–based weather forecasts in an pperational-like context.
\newblock {\em Bulletin of the American Meteorological Society}, 105(6):E864--E883.

\bibitem[Berger and Smith, 2019]{berger_statistical_2019}
Berger, J.~O. and Smith, L.~A. (2019).
\newblock On the statistical formalism of uncertainty quantification.
\newblock {\em Annual Review of Statistics and Its Application}, 6(1):433--460.

\bibitem[Boffetta, 2002]{boffetta_predictability_2002}
Boffetta, G. (2002).
\newblock Predictability: {A} way to characterize complexity.
\newblock {\em Physics Reports}, 356(6):367--474.

\bibitem[Bogner et~al., 2022]{bogner_tercile_2022}
Bogner, K., Chang, A. Y.-Y., Bernhard, L., Zappa, M., Monhart, S., and Spirig, C. (2022).
\newblock Tercile forecasts for extending the horizon of skillful hydrological predictions.
\newblock {\em Journal of Hydrometeorology}, 23(4):521--539.

\bibitem[Bonavita, 2024]{bonavita_limitations_2024}
Bonavita, M. (2024).
\newblock On some limitations of current machine learning weather prediction models.
\newblock {\em Geophysical Research Letters}, 51(12):e2023GL107377.

\bibitem[Boussetta et~al., 2021]{boussetta_ecland_2021}
Boussetta, S., Balsamo, G., Arduini, G., Dutra, E., McNorton, J., Choulga, M., Agustí-Panareda, A., Beljaars, A., Wedi, N., Munõz-Sabater, J., De~Rosnay, P., Sandu, I., Hadade, I., Carver, G., Mazzetti, C., Prudhomme, C., Yamazaki, D., and Zsoter, E. (2021).
\newblock {ECLand}: the {ECMWF} land surface modelling system.
\newblock {\em Atmosphere}, 12(6):723.

\bibitem[Buchovecky et~al., 2023]{buchovecky_potential_2023}
Buchovecky, B., MacGilchrist, G.~A., Bushuk, M., Haumann, F.~A., Frölicher, T.~L., Le~Grix, N., and Dunne, J. (2023).
\newblock Potential predictability of the spring bloom in the southern ocean sea ice zone.
\newblock {\em Geophysical Research Letters}, 50(20):e2023GL105139.

\bibitem[Buizza and Leutbecher, 2015]{buizza_forecast_2015}
Buizza, R. and Leutbecher, M. (2015).
\newblock The forecast skill horizon.
\newblock {\em Quarterly Journal of the Royal Meteorological Society}, 141(693):3366--3382.

\bibitem[Clark et~al., 2001]{clark_ecological_2001}
Clark, J.~S., Carpenter, S.~R., Barber, M., Collins, S., Dobson, A., Foley, J.~A., Lodge, D.~M., Pascual, M., Pielke, R., Pizer, W., Pringle, C., Reid, W.~V., Rose, K.~A., Sala, O., Schlesinger, W.~H., Wall, D.~H., and Wear, D. (2001).
\newblock Ecological forecasts: {An} emerging imperative.
\newblock {\em Science}, 293(5530):657--660.

\bibitem[Dalcher and Kalnay, 1987]{dalcher_error_1987}
Dalcher, A. and Kalnay, E. (1987).
\newblock Error growth and predictability in operational {ECMWF} forecasts.
\newblock {\em Tellus A}, 39A(5):474--491.

\bibitem[Daugaard et~al., 2022]{daugaard_forecasting_2022}
Daugaard, U., Munch, S.~B., Inauen, D., Pennekamp, F., and Petchey, O.~L. (2022).
\newblock Forecasting in the face of ecological complexity: {Number} and strength of species interactions determine forecast skill in ecological communities.
\newblock {\em Ecology Letters}, 25(9):1974--1985.

\bibitem[DelSole and Tippett, 2022]{delsole_statistical_2022}
DelSole, T. and Tippett, M. (2022).
\newblock {\em Statistical methods for climate scientists}.
\newblock Cambridge University Press, 1 edition.

\bibitem[DelSole and Tippett, 2007]{delsole_predictability_2007}
DelSole, T. and Tippett, M.~K. (2007).
\newblock Predictability: {Recent} insights from information theory.
\newblock {\em Reviews of Geophysics}, 45(4).

\bibitem[DelSole and Tippett, 2018]{delsole_predictability_2018}
DelSole, T. and Tippett, M.~K. (2018).
\newblock Predictability in a changing climate.
\newblock {\em Climate Dynamics}, 51(1-2):531--545.

\bibitem[Dietze et~al., 2024]{dietze_near-term_2024}
Dietze, M., White, E.~P., Abeyta, A., Boettiger, C., Bueno~Watts, N., Carey, C.~C., Chaplin-Kramer, R., Emanuel, R.~E., Ernest, S. K.~M., Figueiredo, R.~J., Gerst, M.~D., Johnson, L.~R., Kenney, M.~A., McLachlan, J.~S., Paschalidis, I.~C., Peters, J.~A., Rollinson, C.~R., Simonis, J., Sullivan-Wiley, K., Thomas, R.~Q., Wardle, G.~M., Willson, A.~M., and Zwart, J. (2024).
\newblock Near-term ecological forecasting for climate change action.
\newblock {\em Nature Climate Change}, 14(12):1236--1244.

\bibitem[Dietze, 2017a]{dietze_ecological_2017}
Dietze, M.~C. (2017a).
\newblock {\em Ecological forecasting}.
\newblock Princeton University Press, Princeton.

\bibitem[Dietze, 2017b]{dietze_prediction_2017}
Dietze, M.~C. (2017b).
\newblock Prediction in ecology: a first‐principles framework.
\newblock {\em Ecological Applications}, 27(7):2048--2060.

\bibitem[Dietze et~al., 2018]{dietze_iterative_2018}
Dietze, M.~C., Fox, A., Beck-Johnson, L.~M., Betancourt, J.~L., Hooten, M.~B., Jarnevich, C.~S., Keitt, T.~H., Kenney, M.~A., Laney, C.~M., Larsen, L.~G., Loescher, H.~W., Lunch, C.~K., Pijanowski, B.~C., Randerson, J.~T., Read, E.~K., Tredennick, A.~T., Vargas, R., Weathers, K.~C., and White, E.~P. (2018).
\newblock Iterative near-term ecological forecasting: {Needs}, opportunities, and challenges.
\newblock {\em Proceedings of the National Academy of Sciences}, 115(7):1424--1432.

\bibitem[Dietze et~al., 2023]{dietze_community_2023}
Dietze, M.~C., Thomas, R.~Q., Peters, J., Boettiger, C., Koren, G., Shiklomanov, A.~N., and Ashander, J. (2023).
\newblock A community convention for ecological forecasting: {Output} files and metadata version 1.0.
\newblock {\em Ecosphere}, 14(11):e4686.

\bibitem[Doblas‐Reyes et~al., 2013]{doblasreyes_seasonal_2013}
Doblas‐Reyes, F.~J., García‐Serrano, J., Lienert, F., Biescas, A.~P., and Rodrigues, L. R.~L. (2013).
\newblock Seasonal climate predictability and forecasting: status and prospects.
\newblock {\em WIREs Climate Change}, 4(4):245--268.

\bibitem[Dorigo et~al., 2021]{dorigo_international_2021}
Dorigo, W., Himmelbauer, I., Aberer, D., Schremmer, L., Petrakovic, I., Zappa, L., Preimesberger, W., Xaver, A., Annor, F., Ardö, J., Baldocchi, D., Bitelli, M., Blöschl, G., Bogena, H., Brocca, L., Calvet, J.-C., Camarero, J.~J., Capello, G., Choi, M., Cosh, M.~C., Van De~Giesen, N., Hajdu, I., Ikonen, J., Jensen, K.~H., Kanniah, K.~D., De~Kat, I., Kirchengast, G., Kumar~Rai, P., Kyrouac, J., Larson, K., Liu, S., Loew, A., Moghaddam, M., Martínez~Fernández, J., Mattar~Bader, C., Morbidelli, R., Musial, J.~P., Osenga, E., Palecki, M.~A., Pellarin, T., Petropoulos, G.~P., Pfeil, I., Powers, J., Robock, A., Rüdiger, C., Rummel, U., Strobel, M., Su, Z., Sullivan, R., Tagesson, T., Varlagin, A., Vreugdenhil, M., Walker, J., Wen, J., Wenger, F., Wigneron, J.~P., Woods, M., Yang, K., Zeng, Y., Zhang, X., Zreda, M., Dietrich, S., Gruber, A., Van~Oevelen, P., Wagner, W., Scipal, K., Drusch, M., and Sabia, R. (2021).
\newblock The {International} {Soil} {Moisture} {Network}: serving {Earth} system science for over a decade.
\newblock {\em Hydrology and Earth System Sciences}, 25(11):5749--5804.

\bibitem[Edwards, 2013]{edwards_vast_2013}
Edwards, P.~N. (2013).
\newblock {\em A vast machine: computer models, climate data, and the politics of global warming}.
\newblock Infrastructures series. The MIT Press, Cambridge, Massachusetts London, England, first paperback edition edition.

\bibitem[Evans et~al., 2013]{evans_predictive_2013}
Evans, M.~R., Bithell, M., Cornell, S.~J., Dall, S. R.~X., Díaz, S., Emmott, S., Ernande, B., Grimm, V., Hodgson, D.~J., Lewis, S.~L., Mace, G.~M., Morecroft, M., Moustakas, A., Murphy, E., Newbold, T., Norris, K.~J., Petchey, O., Smith, M., Travis, J. M.~J., and Benton, T.~G. (2013).
\newblock Predictive systems ecology.
\newblock {\em Proceedings of the Royal Society B: Biological Sciences}, 280(1771):20131452.

\bibitem[Fairbairn et~al., 2019]{fairbairn_new_2019}
Fairbairn, D., De~Rosnay, P., and Browne, P.~A. (2019).
\newblock The new stand-alone surface analysis at {ECMWF}: {Implications} for land–atmosphere {DA} coupling.
\newblock {\em Journal of Hydrometeorology}, 20(10):2023--2042.

\bibitem[Frölicher et~al., 2020]{frolicher_potential_2020}
Frölicher, T.~L., Ramseyer, L., Raible, C.~C., Rodgers, K.~B., and Dunne, J. (2020).
\newblock Potential predictability of marine ecosystem drivers.
\newblock {\em Biogeosciences}, 17(7):2061--2083.

\bibitem[Gal and Ghahramani, 2015]{gal_dropout_2015}
Gal, Y. and Ghahramani, Z. (2015).
\newblock Dropout as a {Bayesian} approximation: {Representing} model uncertainty in deep learning.
\newblock Version Number: 6.

\bibitem[Getz et~al., 2018]{getz_making_2018}
Getz, W.~M., Marshall, C.~R., Carlson, C.~J., Giuggioli, L., Ryan, S.~J., Romañach, S.~S., Boettiger, C., Chamberlain, S.~D., Larsen, L., D’Odorico, P., and O’Sullivan, D. (2018).
\newblock Making ecological models adequate.
\newblock {\em Ecology Letters}, 21(2):153--166.

\bibitem[Gneiting and Raftery, 2007]{gneiting_strictly_2007}
Gneiting, T. and Raftery, A.~E. (2007).
\newblock Strictly proper scoring rules, prediction, and estimation.
\newblock {\em Journal of the American Statistical Association}, 102(477):359--378.

\bibitem[Gneiting and Ranjan, 2011]{gneiting_comparing_2011}
Gneiting, T. and Ranjan, R. (2011).
\newblock Comparing density forecasts using threshold- and quantile-weighted scoring rules.
\newblock {\em Journal of Business \& Economic Statistics}, 29(3):411--422.

\bibitem[Guo et~al., 2012]{guo_rebound_2012}
Guo, Z., Dirmeyer, P.~A., DelSole, T., and Koster, R.~D. (2012).
\newblock Rebound in atmospheric predictability and the role of the land surface.
\newblock {\em Journal of Climate}, 25(13):4744--4749.

\bibitem[Hersbach, 2000]{hersbach_decomposition_2000}
Hersbach, H. (2000).
\newblock Decomposition of the continuous ranked probability score for ensemble prediction systems.
\newblock {\em Weather and Forecasting}, 15(5):559--570.

\bibitem[Hersbach et~al., 2020]{hersbach_era5_2020}
Hersbach, H., Bell, B., Berrisford, P., Hirahara, S., Horányi, A., Muñoz‐Sabater, J., Nicolas, J., Peubey, C., Radu, R., Schepers, D., Simmons, A., Soci, C., Abdalla, S., Abellan, X., Balsamo, G., Bechtold, P., Biavati, G., Bidlot, J., Bonavita, M., De~Chiara, G., Dahlgren, P., Dee, D., Diamantakis, M., Dragani, R., Flemming, J., Forbes, R., Fuentes, M., Geer, A., Haimberger, L., Healy, S., Hogan, R.~J., Hólm, E., Janisková, M., Keeley, S., Laloyaux, P., Lopez, P., Lupu, C., Radnoti, G., De~Rosnay, P., Rozum, I., Vamborg, F., Villaume, S., and Thépaut, J. (2020).
\newblock The {ERA5} global reanalysis.
\newblock {\em Quarterly Journal of the Royal Meteorological Society}, 146(730):1999--2049.

\bibitem[Hopson, 2014]{hopson_assessing_2014}
Hopson, T.~M. (2014).
\newblock Assessing the ensemble spread–error relationship.
\newblock {\em Monthly Weather Review}, 142(3):1125--1142.

\bibitem[Hunt et~al., 2004]{hunt_theory_2004}
Hunt, B.~R., Li, T.-Y., Kennedy, J.~A., and Nusse, H.~E., editors (2004).
\newblock {\em The theory of chaotic attractors}.
\newblock Springer New York, New York, NY.

\bibitem[Jolliffe, 2012]{jolliffe_forecast_2012}
Jolliffe, I.~T., editor (2012).
\newblock {\em Forecast verification: a practitioner's guide in atmospheric science}.
\newblock Wiley-Blackwell, Oxford, 2. ed edition.

\bibitem[Jolliffe and Stephenson, 2003]{jolliffe_forecast_2003}
Jolliffe, I.~T. and Stephenson, D.~B., editors (2003).
\newblock {\em Forecast verification: a practitioner's guide in atmospheric science}.
\newblock J. Wiley, Chichester, West Sussex, England ; Hoboken, NJ.

\bibitem[Judd et~al., 2008]{judd_geometry_2008}
Judd, K., Reynolds, C.~A., Rosmond, T.~E., and Smith, L.~A. (2008).
\newblock The geometry of model error.
\newblock {\em Journal of the Atmospheric Sciences}, 65(6):1749--1772.

\bibitem[Jørgensen, 2009]{jorgensen_ecosystem_2009}
Jørgensen, S.~E., editor (2009).
\newblock {\em Ecosystem ecology}.
\newblock Elsevier, Amsterdam, Netherlands ; Boston [Mass.], 1st ed edition.
\newblock OCLC: ocn426812909.

\bibitem[Karunarathna et~al., 2024]{karunarathna_modelling_2024}
Karunarathna, K., Wells, K., and Clark, N.~J. (2024).
\newblock Modelling nonlinear responses of a desert rodent species to environmental change with hierarchical dynamic generalized additive models.
\newblock {\em Ecological Modelling}, 490:110648.

\bibitem[Kazimirović et~al., 2024]{kazimirovic_dynamic_2024}
Kazimirović, M., Stajić, B., Petrović, N., Ljubičić, J., Košanin, O., Hanewinkel, M., and Sperlich, D. (2024).
\newblock Dynamic height growth models for highly productive pedunculate oak ({Quercus} robur {L}.) stands: explicit mapping of site index classification in {Serbia}.
\newblock {\em Annals of Forest Science}, 81(1):15.

\bibitem[Keisler, 2022]{keisler_forecasting_2022}
Keisler, R. (2022).
\newblock Forecasting global weather with graph neural networks.
\newblock arXiv:2202.07575 [physics].

\bibitem[Kramer and Akça, 2008]{kramer_leitfaden_2008}
Kramer, H. and Akça, A. (2008).
\newblock {\em Leitfaden zur {Waldmesslehre}}.
\newblock Sauerländer, Frankfurt, M, 5., überarb. aufl edition.

\bibitem[Kändler and Cullmann, 2014]{kandler_wald_2014}
Kändler, G. and Cullmann, D. (2014).
\newblock Der {Wald} in {Baden}-{Württemberg}.

\bibitem[Landsberg and Waring, 1997]{landsberg_generalised_1997}
Landsberg, J. and Waring, R. (1997).
\newblock A generalised model of forest productivity using simplified concepts of radiation-use efficiency, carbon balance and partitioning.
\newblock {\em Forest Ecology and Management}, 95(3):209--228.

\bibitem[Lewis et~al., 2022]{lewis_power_2022}
Lewis, A. S.~L., Rollinson, C.~R., Allyn, A.~J., Ashander, J., Brodie, S., Brookson, C.~B., Collins, E., Dietze, M.~C., Gallinat, A.~S., Juvigny‐Khenafou, N., Koren, G., McGlinn, D.~J., Moustahfid, H., Peters, J.~A., Record, N.~R., Robbins, C.~J., Tonkin, J., and Wardle, G.~M. (2022).
\newblock The power of forecasts to advance ecological theory.
\newblock {\em Methods in Ecology and Evolution}, pages 2041--210X.13955.

\bibitem[Lorenz, 1963]{lorenz_deterministic_1963}
Lorenz, E.~N. (1963).
\newblock Deterministic {Nonperiodic} {Flow}.
\newblock {\em Journal of the Atmospheric Sciences}, 20.

\bibitem[Lorenz, 1982]{lorenz_atmospheric_1982}
Lorenz, E.~N. (1982).
\newblock Atmospheric predictability experiments with a large numerical model.
\newblock {\em Tellus A: Dynamic Meteorology and Oceanography}, 34(6):505.

\bibitem[Lorenz, 1996]{lorenz_predictability_1996}
Lorenz, E.~N. (1996).
\newblock Predictability - {A} problem partly solved.

\bibitem[Magnusson et~al., 2019]{magnusson_dependence_2019}
Magnusson, L., Chen, J., Lin, S., Zhou, L., and Chen, X. (2019).
\newblock Dependence on initial conditions versus model formulations for medium‐range forecast error variations.
\newblock {\em Quarterly Journal of the Royal Meteorological Society}, 145(722):2085--2100.

\bibitem[Massoud et~al., 2018]{massoud_probing_2018}
Massoud, E.~C., Huisman, J., Benincà, E., Dietze, M.~C., Bouten, W., and Vrugt, J.~A. (2018).
\newblock Probing the limits of predictability: data assimilation of chaotic dynamics in complex food webs.
\newblock {\em Ecology Letters}, 21(1):93--103.

\bibitem[May, 1974]{may_biological_1974}
May, R.~M. (1974).
\newblock Biological populations with nonoverlapping generations: {Stable} points, stable cycles, and chaos.
\newblock {\em Science}, 186(4164):645--647.

\bibitem[McWilliams, 2018]{mcwilliams_perspective_2018}
McWilliams, J.~C. (2018).
\newblock A perspective on the legacy of {Edward} {Lorenz}.
\newblock {\em Earth and Space Science}, (6):336--350.

\bibitem[Msadek et~al., 2010]{msadek_assessing_2010}
Msadek, R., Dixon, K.~W., Delworth, T.~L., and Hurlin, W. (2010).
\newblock Assessing the predictability of the {Atlantic} meridional overturning circulation and associated fingerprints: {Assessing} the predictability of the {AMOC}.
\newblock {\em Geophysical Research Letters}, 37(19):n/a--n/a.

\bibitem[Murphy, 1993]{murphy_what_1993}
Murphy, A.~H. (1993).
\newblock What is a good forecast? {An} essay on the nature of goodness in weather forecasting.
\newblock {\em Weather and Forecasting}, 8(2):281--293.

\bibitem[Nearing et~al., 2024]{nearing_global_2024}
Nearing, G., Cohen, D., Dube, V., Gauch, M., Gilon, O., Harrigan, S., Hassidim, A., Klotz, D., Kratzert, F., Metzger, A., Nevo, S., Pappenberger, F., Prudhomme, C., Shalev, G., Shenzis, S., Tekalign, T.~Y., Weitzner, D., and Matias, Y. (2024).
\newblock Global prediction of extreme floods in ungauged watersheds.
\newblock {\em Nature}, 627(8004):559--563.

\bibitem[Owens and Hewson, 2018]{owens_ecmwf_2018}
Owens, R.~G. and Hewson, T.~D. (2018).
\newblock {ECMWF} {Forecast} {User} {Guide}.
\newblock doi: 10.21957/m1cs7h.

\bibitem[Palamara et~al., 2016]{palamara_effects_2016}
Palamara, G.~M., Carrara, F., Smith, M.~J., and Petchey, O.~L. (2016).
\newblock The effects of demographic stochasticity and parameter uncertainty on predicting the establishment of introduced species.
\newblock {\em Ecology and Evolution}, 6(23):8440--8451.

\bibitem[Pappenberger et~al., 2015]{pappenberger_how_2015}
Pappenberger, F., Ramos, M., Cloke, H., Wetterhall, F., Alfieri, L., Bogner, K., Mueller, A., and Salamon, P. (2015).
\newblock How do {I} know if my forecasts are better? {Using} benchmarks in hydrological ensemble prediction.
\newblock {\em Journal of Hydrology}, 522:697--713.

\bibitem[Pennekamp et~al., 2019]{pennekamp_intrinsic_2019}
Pennekamp, F., Iles, A.~C., Garland, J., Brennan, G., Brose, U., Gaedke, U., Jacob, U., Kratina, P., Matthews, B., Munch, S., Novak, M., Palamara, G.~M., Rall, B.~C., Rosenbaum, B., Tabi, A., Ward, C., Williams, R., Ye, H., and Petchey, O.~L. (2019).
\newblock The intrinsic predictability of ecological time series and its potential to guide forecasting.
\newblock {\em Ecological Monographs}, 89(2).

\bibitem[Petchey et~al., 2015]{petchey_ecological_2015}
Petchey, O.~L., Pontarp, M., Massie, T.~M., Kéfi, S., Ozgul, A., Weilenmann, M., Palamara, G.~M., Altermatt, F., Matthews, B., Levine, J.~M., Childs, D.~Z., McGill, B.~J., Schaepman, M.~E., Schmid, B., Spaak, P., Beckerman, A.~P., Pennekamp, F., and Pearse, I.~S. (2015).
\newblock The ecological forecast horizon, and examples of its uses and determinants.
\newblock {\em Ecology Letters}, 18(7):597--611.

\bibitem[Poggio et~al., 2021]{poggio_soilgrids_2021}
Poggio, L., De~Sousa, L.~M., Batjes, N.~H., Heuvelink, G. B.~M., Kempen, B., Ribeiro, E., and Rossiter, D. (2021).
\newblock {SoilGrids} 2.0: producing soil information for the globe with quantified spatial uncertainty.
\newblock {\em SOIL}, 7(1):217--240.

\bibitem[Raiho et~al., 2020]{raiho_towards_2020}
Raiho, A., Dietze, M., Dawson, A., Rollinson, C.~R., Tipton, J., and McLachlan, J. (2020).
\newblock Towards understanding predictability in ecology: {A} forest gap model case study.
\newblock preprint, Ecology.

\bibitem[Rammer and Seidl, 2015]{rammer_coupling_2015}
Rammer, W. and Seidl, R. (2015).
\newblock Coupling human and natural systems: {Simulating} adaptive management agents in dynamically changing forest landscapes.
\newblock {\em Global Environmental Change}, 35:475--485.

\bibitem[Rammer et~al., 2024]{rammer_individual-based_2024}
Rammer, W., Thom, D., Baumann, M., Braziunas, K., Dollinger, C., Kerber, J., Mohr, J., and Seidl, R. (2024).
\newblock The individual-based forest landscape and disturbance model {iLand}: {Overview}, progress, and outlook.
\newblock {\em Ecological Modelling}, 495:110785.

\bibitem[Razafimaharo et~al., 2020]{razafimaharo_new_2020}
Razafimaharo, C., Krähenmann, S., Höpp, S., Rauthe, M., and Deutschländer, T. (2020).
\newblock New high-resolution gridded dataset of daily mean, minimum, and maximum temperature and relative humidity for {Central} {Europe} ({HYRAS}).
\newblock {\em Theoretical and Applied Climatology}, 142(3):1531--1553.

\bibitem[Record et~al., 2023]{record_synthesizing_2023}
Record, S., Boettiger, C., and Rollinson, C.~R. (2023).
\newblock Synthesizing forecasts to inform decision‐making and advance ecological theory.
\newblock {\em Methods in Ecology and Evolution}, 14(3):728--731.

\bibitem[Reggiani et~al., 2024]{reggiani_time-horizons_2024}
Reggiani, P., Biondi, D., and Todini, E. (2024).
\newblock On time-horizons based post-processing of flow forecasts.
\newblock {\em Frontiers in Water}, 6:1359750.

\bibitem[Ricker, 1954]{ricker_stock_1954}
Ricker, W.~E. (1954).
\newblock Stock and {Recruitment}.
\newblock {\em Journal of the Fisheries Research Board of Canada}, 11(5):559--623.

\bibitem[Rogers et~al., 2021]{rogers_chaos_2021}
Rogers, T., Johnson, B., and Munch, S. (2021).
\newblock Chaos is not rare in natural ecosystems.
\newblock preprint, In Review.

\bibitem[Schaeffer et~al., 2024]{schaeffer_forecasting_2024}
Schaeffer, B.~A., Reynolds, N., Ferriby, H., Salls, W., Smith, D., Johnston, J.~M., and Myer, M. (2024).
\newblock Forecasting freshwater cyanobacterial harmful algal blooms for {Sentinel}-3 satellite resolved {U}.{S}. lakes and reservoirs.
\newblock {\em Journal of Environmental Management}, 349:119518.

\bibitem[Scheffer et~al., 2001]{scheffer_catastrophic_2001}
Scheffer, M., Carpenter, S., Foley, J.~A., Folke, C., and Walker, B. (2001).
\newblock Catastrophic shifts in ecosystems.
\newblock {\em Nature}, 413(6856):591--596.

\bibitem[Seidl et~al., 2014]{seidl_simulating_2014}
Seidl, R., Rammer, W., and Blennow, K. (2014).
\newblock Simulating wind disturbance impacts on forest landscapes: {Tree}-level heterogeneity matters.
\newblock {\em Environmental Modelling \& Software}, 51:1--11.

\bibitem[Seidl et~al., 2012a]{seidl_individual-based_2012}
Seidl, R., Rammer, W., Scheller, R.~M., and Spies, T.~A. (2012a).
\newblock An individual-based process model to simulate landscape-scale forest ecosystem dynamics.
\newblock {\em Ecological Modelling}, 231:87--100.

\bibitem[Seidl et~al., 2012b]{seidl_multi-scale_2012}
Seidl, R., Spies, T.~A., Rammer, W., Steel, E.~A., Pabst, R.~J., and Olsen, K. (2012b).
\newblock Multi-scale drivers of spatial variation in old-growth forest carbon density disentangled with lidar and an individual-based landscape model.
\newblock {\em Ecosystems}, 15(8):1321--1335.

\bibitem[Shen et~al., 2023]{shen_lorenzs_2023}
Shen, B.-W., Pielke, R.~A., Zeng, X., and Zeng, X. (2023).
\newblock Lorenz’s view on the predictability limit of the atmosphere.
\newblock {\em Encyclopedia}, 3(3):887--899.

\bibitem[Smith, 2006]{palmer_predictability_2006}
Smith, L.~A. (2006).
\newblock Predictability past, predictability present.
\newblock In Palmer, T. and Hagedorn, R., editors, {\em Predictability of {Weather} and {Climate}}, pages 217--250. Cambridge University Press, 1 edition.

\bibitem[Smith et~al., 1999]{smith_uncertainty_1999}
Smith, L.~A., Ziehmann, C., and Fraedrich, K. (1999).
\newblock Uncertainty dynamics and predictability in chaotic systems.
\newblock {\em Quarterly Journal of the Royal Meteorological Society}, 125(560):2855--2886.

\bibitem[Spring and Ilyina, 2020]{spring_predictability_2020}
Spring, A. and Ilyina, T. (2020).
\newblock Predictability horizons in the global carbon cycle inferred from a perfect‐model framework.
\newblock {\em Geophysical Research Letters}, 47(9).

\bibitem[Subbey et~al., 2014]{subbey_modelling_2014}
Subbey, S., Devine, J.~A., Schaarschmidt, U., and Nash, R.~D. (2014).
\newblock Modelling and forecasting stock–recruitment: current and future perspectives.
\newblock {\em ICES Journal of Marine Science}, 71(8):2307--2322.

\bibitem[Sun and Zhang, 2016]{sun_intrinsic_2016}
Sun, Y.~Q. and Zhang, F. (2016).
\newblock Intrinsic versus practical limits of atmospheric predictability and the significance of the butterfly effect.
\newblock {\em Journal of the Atmospheric Sciences}, 73(3):1419--1438.

\bibitem[Séférian et~al., 2018]{seferian_assessing_2018}
Séférian, R., Berthet, S., and Chevallier, M. (2018).
\newblock Assessing the decadal predictability of land and ocean carbon uptake.
\newblock {\em Geophysical Research Letters}, 45(5):2455--2466.

\bibitem[Thom et~al., 2024]{thom_parameters_2024}
Thom, D., Rammer, W., Albrich, K., Braziunas, K.~H., Dobor, L., Dollinger, C., Hansen, W.~D., Harvey, B.~J., Hlásny, T., Hoecker, T.~J., Honkaniemi, J., Keeton, W.~S., Kobayashi, Y., Kruszka, S.~S., Mori, A., Morris, J.~E., Peters-Collaer, S., Ratajczak, Z., Simensen, T., Storms, I., Suzuki, K.~F., Taylor, A.~R., Turner, M.~G., Willis, S., and Seidl, R. (2024).
\newblock Parameters of 150 temperate and boreal tree species and provenances for an individual-based forest landscape and disturbance model.
\newblock {\em Data in Brief}, 55:110662.

\bibitem[Thom et~al., 2017]{thom_impacts_2017}
Thom, D., Rammer, W., Dirnböck, T., Müller, J., Kobler, J., Katzensteiner, K., Helm, N., and Seidl, R. (2017).
\newblock The impacts of climate change and disturbance on spatio‐temporal trajectories of biodiversity in a temperate forest landscape.
\newblock {\em Journal of Applied Ecology}, 54(1):28--38.

\bibitem[Thom et~al., 2022]{thom_will_2022}
Thom, D., Rammer, W., Laux, P., Smiatek, G., Kunstmann, H., Seibold, S., and Seidl, R. (2022).
\newblock Will forest dynamics continue to accelerate throughout the 21st century in the {Northern} {Alps}?
\newblock {\em Global Change Biology}, 28(10):3260--3274.

\bibitem[Thomas et~al., 2023]{thomas_span_2023}
Thomas, R.~Q., Boettiger, C., Carey, C.~C., Dietze, M.~C., Johnson, L.~R., Kenney, M.~A., McLachlan, J.~S., Peters, J.~A., Sokol, E.~R., Weltzin, J.~F., Willson, A., Woelmer, W.~M., and {Challenge contributors} (2023).
\newblock The {\textless}span style="font-variant:small-caps;"{\textgreater}{NEON}{\textless}/span{\textgreater} {Ecological} {Forecasting} {Challenge}.
\newblock {\em Frontiers in Ecology and the Environment}, 21(3):112--113.

\bibitem[Thomas et~al., 2020]{thomas_nearterm_2020}
Thomas, R.~Q., Figueiredo, R.~J., Daneshmand, V., Bookout, B.~J., Puckett, L.~K., and Carey, C.~C. (2020).
\newblock A near‐term iterative forecasting system successfully predicts reservoir hydrodynamics and partitions uncertainty in real time.
\newblock {\em Water Resources Research}, 56(11).

\bibitem[Tiedje et~al., 2012]{tiedje_potential_2012}
Tiedje, B., Köhl, A., and Baehr, J. (2012).
\newblock Potential predictability of the north atlantic heat transport based on an oceanic state estimate.
\newblock {\em Journal of Climate}, 25(24):8475--8486.

\bibitem[Urban et~al., 2022]{urban_coding_2022}
Urban, M.~C., Travis, J. M.~J., Zurell, D., Thompson, P.~L., Synes, N.~W., Scarpa, A., Peres-Neto, P.~R., Malchow, A.-K., James, P. M.~A., Gravel, D., De~Meester, L., Brown, C., Bocedi, G., Albert, C.~H., Gonzalez, A., and Hendry, A.~P. (2022).
\newblock Coding for life: {Designing} a platform for projecting and protecting global biodiversity.
\newblock {\em BioScience}, 72(1):91--104.

\bibitem[Wesselkamp et~al., 2025]{wesselkamp_advances_2025}
Wesselkamp, M., Chantry, M., Pinnington, E., Choulga, M., Boussetta, S., Kalweit, M., Bödecker, J., Dormann, C.~F., Pappenberger, F., and Balsamo, G. (2025).
\newblock Advances in land surface forecasting: a comparison of {LSTM}, gradient boosting, and feed-forward neural networks as prognostic state emulators in a case study with {ecLand}.
\newblock {\em Geoscientific Model Development}, 18(4):921--937.

\bibitem[Wheeler et~al., 2024]{wheeler_predicting_2024}
Wheeler, K.~I., Dietze, M.~C., LeBauer, D., Peters, J.~A., Richardson, A.~D., Ross, A.~A., Thomas, R.~Q., Zhu, K., Bhat, U., Munch, S., Floreani~Buzbee, R., Chen, M., Goldstein, B., Guo, J., Hao, D., Jones, C., Kelly-Fair, M., Liu, H., Malmborg, C., Neupane, N., Pal, D., Shirey, V., Song, Y., Steen, M., Vance, E.~A., Woelmer, W.~M., Wynne, J.~H., and Zachmann, L. (2024).
\newblock Predicting spring phenology in deciduous broadleaf forests: {NEON} phenology forecasting community challenge.
\newblock {\em Agricultural and Forest Meteorology}, 345:109810.

\bibitem[Woelmer et~al., 2022]{woelmer_nearterm_2022}
Woelmer, W.~M., Thomas, R.~Q., Lofton, M.~E., McClure, R.~P., Wander, H.~L., and Carey, C.~C. (2022).
\newblock Near‐term phytoplankton forecasts reveal the effects of model time step and forecast horizon on predictability.
\newblock {\em Ecological Applications}, 32(7).

\bibitem[Zhang et~al., 2019a]{zhang_what_2019}
Zhang, F., Sun, Y.~Q., Magnusson, L., Buizza, R., Lin, S.-J., Chen, J.-H., and Emanuel, K. (2019a).
\newblock What is the predictability limit of midlatitude weather?
\newblock {\em Journal of the Atmospheric Sciences}, 76(4):1077--1091.

\bibitem[Zhang et~al., 2019b]{zhang_identification_2019}
Zhang, S., Meurey, C., and Calvet, J.-C. (2019b).
\newblock Identification of soil-cooling rains in southern {France} from soil temperature and soil moisture observations.
\newblock {\em Atmospheric Chemistry and Physics}, 19(7):5005--5020.

\bibitem[Zhou et~al., 2024]{zhou_evaluation_2024}
Zhou, J., Zhang, J., and Huang, Y. (2024).
\newblock Evaluation of soil temperature in {CMIP6} multimodel simulations.
\newblock {\em Agricultural and Forest Meteorology}, 352:110039.

\bibitem[Zwart et~al., 2023]{zwart_nearterm_2023}
Zwart, J.~A., Oliver, S.~K., Watkins, W.~D., Sadler, J.~M., Appling, A.~P., Corson‐Dosch, H.~R., Jia, X., Kumar, V., and Read, J.~S. (2023).
\newblock Near‐term forecasts of stream temperature using deep learning and data assimilation in support of management decisions.
\newblock {\em JAWRA Journal of the American Water Resources Association}, 59(2):317--337.

\end{thebibliography}

\clearpage
\appendix

\section{Case study 1 : Stochastic Ricker equation}

\subsection{Model description} Ricker equations are used in theoretical ecology to describe functional responses of density-dependent population growth in community structures \citep{ricker_stock_1954, subbey_modelling_2014}. In a coupled version they constitute the simplest description of an ecosystem, containing at least one term of interaction with other units, i.e. populations, of the system \citep{jorgensen_ecosystem_2009}. Under certain parameter settings, the Ricker equation becomes unstable, resulting in chaotic system behaviour \citep{may_biological_1974}. 
The Ricker equation is a time-discrete function and for one system component iteratively defined as 
\begin{equation}\label{ricker}
y_{t} = y_{t-1} e^{\alpha(1-\beta y_{t-1})}, 
\end{equation}
where $\alpha$ is the density independent growth rate and $\beta\cdot y_{t-1}$ the density dependent term. The state variable at any time $t = 1, ..., \tau$ is the population size $y_t$ which we define relatively to the systems carrying capacity that hence standardises to 1.

\paragraph{Stochastic Ricker equation} An ensemble of 500 trajectories was simulated by sampling the parameters of the Ricker equation at each time step from uni-variate normal distributions with a variance of $\epsilon = 10\%$ of the parameter size. The $\beta$ in equation \ref{ricker} can also be written as $\frac{1}{k}$ where $k$ is the carrying capacity. While we set $k = 1$ for the coupled Ricker equation (see below), here we simulated the carrying capacity also stochastically. 
\begin{align}
y_0 &= 1 + \sigma_{y_0} \\
r &\sim \mathcal{N}(0.1, 0.1\epsilon) \\
k &\sim \mathcal{N}(2, 2\epsilon) \\
\sigma_{y_0} &\sim \mathcal{N}(0.01, 0.01\epsilon) \\
\end{align}

\section{Case study 2: iLand}

\subsection{Model and variable description}

We applied the individual-based forest landscape and disturbance model iLand to demonstrate the actual forecast limit (see section 3.2) on the example of simulated tree productivity. iLand is a high-resolution, process-based model simulating forest ecosystem processes on multiple scales, from individual trees to landscapes \citep{seidl_individual-based_2012}. Here, we only describe model components of particular relevance for simulating tree productivity in managed monospecific stands. Primary production is simulated using a resource-use efficiency model \citep{landsberg_generalised_1997} and is driven by daily resolved climate variables (temperature, precipitation, radiation and vapour pressure deficit), atmospheric CO2 concentration and stable soil conditions (sand, silt and clay fractions, effective soil depth and available nitrogen). Carbohydrate acquisition for each tree is determined by its competitive position for available light. Tree mortality is influenced by the trees age and size as well as its carbon balance (stress-related mortality). Growth, survival, and regeneration in iLand are controlled by 61 species-specific model parameters \citep{thom_parameters_2024}, allowing for the simulation of species' unique responses to environmental changes . iLand was previously parameterised, evaluated and applied for Central European ecosystems \citep{albrich_tradeoffs_2018, thom_impacts_2017, thom_will_2022, seidl_simulating_2014} and we used the default model parameters from the iLand 1.0 example landscape. Management interventions were implemented using the agent-based forest management model ABE \citep{rammer_coupling_2015}, which is fully integrated into the iLand simulation framework. A detailed description of iLand can be found in \citet{seidl_individual-based_2012, seidl_multi-scale_2012} and \citet{rammer_individual-based_2024}. iLand model code, software and documentation are available online via \url{http://ilandmodel.org/}.

\subsection{Model initialisation and parameterisation}


To initialise and simulate the one-hectare monospecific test stands within the Freiburger Stadtwald, information on soils, current vegetation and daily climate was needed (see Table \ref{tab:iLand_datasources} for an overview of the variables needed and data sources used). Based on the fine-tuning of a previous study in the Black Forest National Park by Kern et al. (unpublished), the values for plant available nitrogen (nav) and epsilon (\(\varepsilon_0\)), the biome-specific optimum Light Use Efficiency (LUE), were set to 100 kg ha\(^{-1}\) a\(^{-1}\) and 2.1 gC MJ\(^{-1}\), respectively. The carbon cycle was disabled for the simulations.

The climate was represented by observed climate for the period 1951-2020. Historical climate data were extracted from the HYRAS-DE raster dataset \citep{razafimaharo_new_2020} and downscaled from 5x5 km to 100x100 m, following \citet{thom_will_2022}. The method is based on the relationship between climatic variables and elevation. A daily laps rate for each climate variable is calculated and then used for downscaling each 100x100 m-cell’s climate parameters based on the cell’s elevation. The CO2 concentration remained constant over the whole simulation period.  

\begin{table}[ht]\label{tab:iLand_datasources}
    \centering
    \caption{Overview of variables needed for model initialisation.}
    \renewcommand{\arraystretch}{1.5}
    \begin{tabular}{|l|l|p{6cm}|}  
        \hline
        \textbf{Variable group} & \textbf{Variables} & \textbf{Data source} \\ \hline
        \multirow{2}{*}{Soil} 
        & Effective soil depth & \multirow{2}{*}{\raggedright \citep{poggio_soilgrids_2021}}\\ \cline{2-2}
        & Soil texture (shares of sand, silt, and clay) & \\ \hline
        \multirow{4}{*}{Climate} 
        & Daily temperatures (Min, Max) [°C]& \multirow{4}{*}{\raggedright \shortstack[1]{HYRAS-DE v5; HYRAS-DE-RSDS v3 \\ 
        \citep{razafimaharo_new_2020}}} \\ \cline{2-2} 
        & Precipitation [mm/day]& \\ \cline{2-2}
        & Solar radiation [MJ/m2/day]& \\ \cline{2-2}
        & Vapour pressure deficit [kPa]& \\ \hline
        \multirow{4}{*}{Vegetation} 
        & Tree count & \multirow{4}{*}{\raggedright \shortstack[1]{Yield tables Baden Württemberg\\ 
        FVA-BW (\url{www.fva-bw.de})}} \\ \cline{2-2}
        & DBH range & \\ \cline{2-2}
        & Height-diameter ratio & \\ \cline{2-2}
        & Age & \\ \hline
    \end{tabular}
\end{table}

\subsection{Simulation}

To evaluate tree productivity in managed monospecific stands as simulated by iLand, we compared predicted and observed dominant height over a 65-year period. Observations included both point measurements and reconstructions based on yield tables. The dominant height (h100) of a stand is defined as the average height of the 100 largest trees. Initially, stands were identified for which the SI100, the dominant height (h100) at age 100, could be determined. For this purpose, the available inventory plots of the third BWI \citep{kandler_wald_2014} were used, which are within the Freiburger Stadtwald (only the “Bergwald” inventory plots were available). Each species measured in a plot is assigned with a SI100 using yield tables, which provide a SI100 estimate based on dominant height and age. Based on the SI100 the dominant height growth of each individual tree is reconstructed from yield tables over the whole period. The SI100 derived from the point measurements (observation) defines the yield class of each measured tree. The dominant height growth of the corresponding yield class described in the yield tables is pinned to the point observation to generate a time series. The inventory plots are taken spatially explicit as input stands for iLand, with each plot represented by a 100x100 meter cell, corresponding to a stand for each species at the respective inventory plot. In total 269 (instead of the original 270, we dropped Pinus Sylvestris for which only one observation was available) testing stands were modelled, each monospecific and with an observed SI100 as well as reconstructed dominant height growth. The vegetation was initialized at an age of 45 years, based on stand variables (see Table \ref{tab:iLand_datasources}) derived from the yield tables. The stand development was then simulated with a thinning management applied in five-year intervals, reducing the number of stems per hectare according to the yield tables. For these simulations, the available reference climate data from 1951 to 2016 was used in chronological order. 

\begin{figure}
    \centering
    \includegraphics[width=0.5\linewidth]{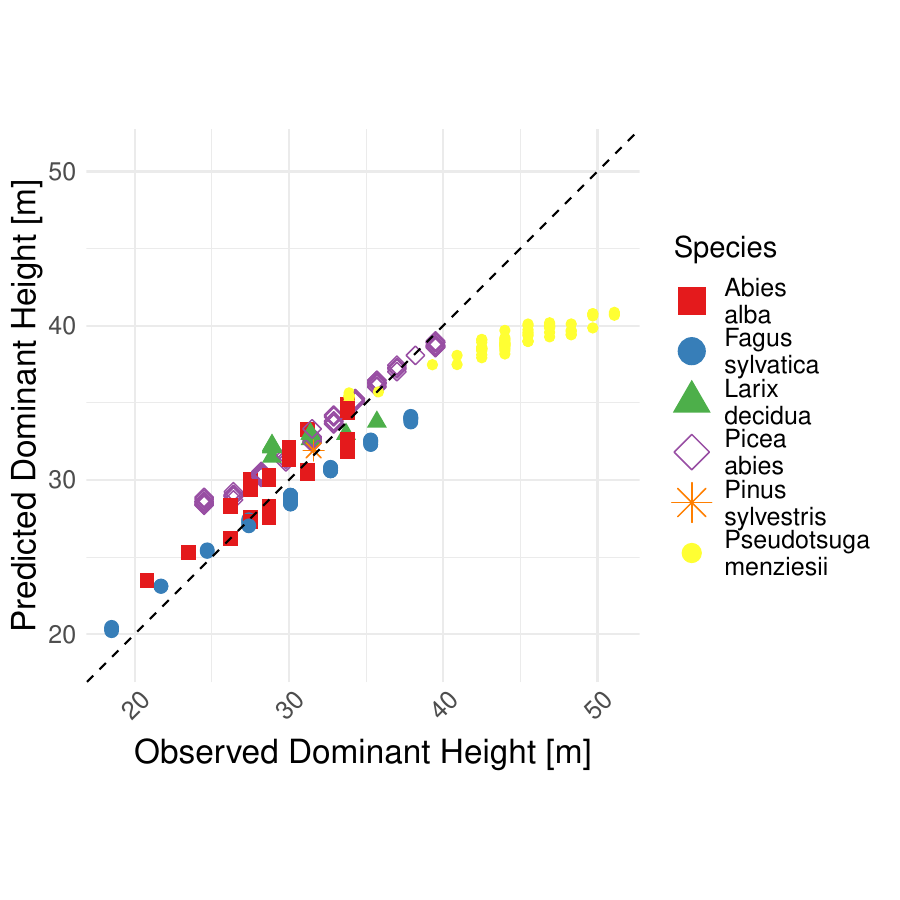}
    \caption{Correlation of observed and simulated dominant heights at age 100.}
    \label{fig:enter-label}
\end{figure}

\subsection{Algorithm for determining the forecast limit}

\begin{algorithm}[H]
\caption{Generalized process to receive $\hat{h}(g)$ for all tree species, where $g$ is the observed yield class of a stand. }
\begin{algorithmic}[1]
\FOR{each species in $\text{tree species}$}
            
    \STATE Get forecast and observations for species
    \STATE $\mathbf{\hat{Y}} \gets \text{dominant\_height\_forecast[species]}$
    \STATE $\mathbf{Y} \gets \text{dominant\_height\_observations[species]}$
    
    \FOR{each stand in $\text{species\_stands}$}
        \STATE Select stand
        \STATE $\mathbf{\hat{y}}(g) \gets \mathbf{\hat{Y}}[\text{idx}]$
        \STATE $\mathbf{y}(g) \gets \mathbf{Y}[\text{idx}]$
        
        \STATE Calculate proficiency
        \STATE $\mathbf{e} \gets |(\mathbf{y}(g) - \mathbf{\hat{y}}(g))|$
        \STATE $\mathbf{e}_{b} \gets |(\mathbf{y}((g \pm \rho_{g}) - \mathbf{\hat{y}}(g))|$

        \STATE Get threshold
        \STATE $\rho \gets \underset{t}{\text{min}}(\mathbf{e} \geq \mathbf{e}_{b})$
        
        \STATE Calculate bounded error trajectories $\mathbf{f}$
        \IF{$\rho$ is not infinite}
            \STATE $\mathbf{f} \gets (\rho - \mathbf{e})$
        \ENDIF

    \ENDFOR
    

\ENDFOR
\end{algorithmic}
\end{algorithm}

\section{Case study 3: aiLand}

\subsection{Model and variable description}

The forecast model $\mathcal{M}(Y_0, X, \theta)$ is a model-ensemble-based machine learning emulator of ECMWF's physical land surface scheme ecLand (hereafter: aiLand) \citep{boussetta_ecland_2021, wesselkamp_advances_2025} which we use to demonstrate both absolute and relative forecast limits \citep{buizza_forecast_2015}. aiLand refers to three models: A feed forward neural network, a long short-term memory neural network and an extreme gradient boosting algorithm \citep{wesselkamp_advances_2025}. They were parametrised on a European scale with 10-years of (historic) numerical simulations at 6-hourly temporal resolution on a 31 km spatial resolution grid. The details on training and test data and on the parametrisation procedure for the machine learning models, as well as the description of according ecLand simulations that are in this work used as the uninitialised physical model reference can all be found in \citet{wesselkamp_advances_2025} and its supplementary material. 
State variables $\widehat{Y}$ are soil water volume (m$^3$m$^{-3}$) and soil temperature (K) at the soil surface layer (0-5 cm), subsurface layer 1 (5-20 cm) and subsurface layer 2 (20-70 cm), and snow cover. All states together initialise $\mathcal{M}$ as $Y_0$. 
External processes $X$ that force $Y$ are ERA-5 reanalysis dynamic meteorological variables and static climate and physiographic fields \citep{hersbach_era5_2020}. More details on model and variables are described in \citet{wesselkamp_advances_2025}.

\subsection{Evaluation with ISMN observations}

aiLand was evaluated with in situ soil temperature and soil moisture data that was pre-processed along \citep{fairbairn_new_2019}. These were assembled from International Soil Moisture Network (ISMN) observations of the french SMOSMANIA network \citep{dorigo_international_2021}. As initial time, the 01.02.22, i.e.~most recent year was chosen, which required an interval of validation data from the lookback time of the LSTM (23.01.2021) up to the medium-range lead time (14.02.2022). For this time period, 16 of 21 stations had complete time series of soil temperature measurements and 8 stations of soil moisture measurements. For the stations we used, their soil types are displayed in table \ref{tab:smosmania_data}. For more information, see \citet{zhang_identification_2019} and \citet{dorigo_international_2021}.

The station data were matched with the according grid cell of the physiographic and climate fields, which were used to force the models during the period. Station data were further resampled to the 6-hourly resolution of the forcing data for evaluation and standardised by z-scoring with the same ecLand prognostic global mean and standard deviation used to train the models in \citet{wesselkamp_advances_2025}. 


\begin{table}[ht]
\centering
\begin{tabular}{|l|l|l|r|r|r|}
\hline
\textbf{Station} & \textbf{Soil Type}  \\
\hline
Condom & Silty clay \\
Villevielle & Sandy loam \\
LaGrandCombe & LaGrandCombe \\
Narbonne & Clay \\
Urgons & Silt loam  \\
LezignanCorbieres & Sandy clay loam  \\
CabrieresdAvignon & Sandy clay loam \\
Savenes & Loam \\
PeyrusseGrande & Silty clay \\
Sabres & Sand \\
Montaut & Montaut  \\
Mazan-Abbaye & Sandy loam \\
Mouthoumet & Clay loam \\
Mejannes-le-Clap & Loam \\
CreondArmagnac & Sand \\
SaintFelixdeLauragais & Loam \\

\hline
\end{tabular}
\caption{French stations from the SMOSMANIA network, used for computing soil temperature limits on three different layers. (\url{https://doi.org/10.5194/acp-19-5005-2019})}
\label{tab:smosmania_data}
\end{table}

\subsection{Deterministic forecast limit}

\paragraph{Experimental setup}

We show our analysis for soil temperature. aiLand is evaluated as a model ensemble on observations from the International Soil Moisture Network (ISMN) that provides a quality-controlled and harmonised data base for land-surface process evaluation \citep{dorigo_international_2021}. We use local observations from years 2021 and 2022 of $m=13$ French Smosmania network stations, that we define as the reference $\textbf{Y}$. Soil temperature was measured at 0-5 cm (surface layer), 5-20 cm (subsurface layer 1) and 20-30 cm (subsurface layer 2). aiLand is exemplarily initialised with observations on February 1st 2022, constituting initial conditions $Y_0$, and is then integrated at a 6-hourly temporal resolution over $\tau = 1, \cdots 56$ lead times, representing the medium-range. For long-range forecasts $\tau = 1, \cdots 1200$ and soil water volume results, see Appendix.

\paragraph{Actual and relative forecast limits} 

We compute an absolute and a relative forecast limit of aiLand. Scoring function $\mathcal{S}$ is the mean absolute error (MAE) and the ad-hoc scoring tolerance for soil temperature is $\varrho = 1.5 K$ \citep{zhou_evaluation_2024}. The relative forecast limit is determined by computing aiLand skill toward the numerical ecLand as null model, as such $\mathcal{S}$ becomes the MAE-SS, a first order skill scoring function \citep{jolliffe_forecast_2012}. The MAE is spatially averaged over stations and, considering aiLand is as a model-ensemble, can be computed as
\begin{equation}
    \text{MAE}_{\text{aiLand}_t} = \frac{1}{m} \sum |\overline{\widehat{Y}}_{t} - \mathbf{Y}_{t,m}|,\quad\text{where}\quad m = 1,\dots,13\quad\text{and}\quad t = 1, \dots, \tau
\end{equation}
where $\overline{\widehat{Y}}_{m}$ is the model-ensemble mean. Skill is hence stepwise evaluated as
\begin{equation}\label{eq:MAESS}
    \text{MAE-SS}_t = 1 - \frac{\text{MAE}_{\text{aiLand}}(\overline{\widehat{Y}}_{t}, \mathbf{Y}_{t,m})}{\text{AE}_{\text{ecLand}}(\widehat{Y}_t, \mathbf{Y}_{t,m})}, \quad\text{where}\quad t = 1, \dots, \tau,
\end{equation}
and where the MAE step-wise averages over the emulator model-ensemble, but in fact, for only one station reduces the point forecast of physical ecLand to a point-wise absolute error.

\paragraph{Results} 

\begin{figure}[h]
    \centering
    \includegraphics[width=0.99\linewidth]{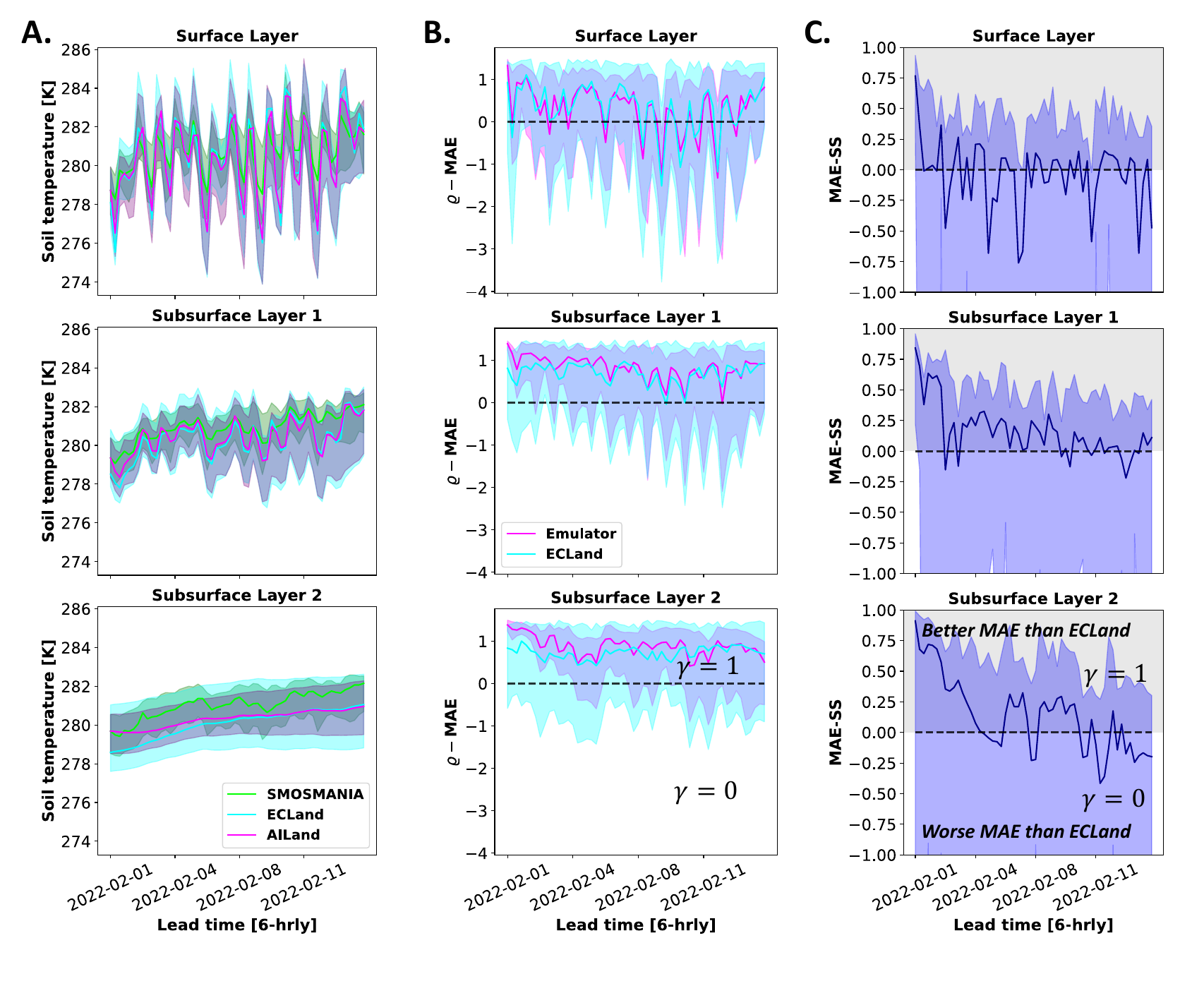}
    \caption{ \textbf{A.} Forecast of aiLand and of numerical ecLand for soil temperature measurements across stations at the surface and two subsurface layers over a medium-range test period in February 2022. \textbf{B.} Actual forecast limits of aiLand and ecLand with a tolerance of 1.5 K toward the predictive MAE on the station measurements. \textbf{C.} Relative forecast limit based on the MAE-SS of aiLand toward ecLand. Gray shaded regions indicate areas where the MAE of aiLand is smaller than that of ecLand.}
    \label{fig:aiLand_appendix_temp}
\end{figure}

\begin{figure}
    \centering
    \includegraphics[width=0.95\linewidth]{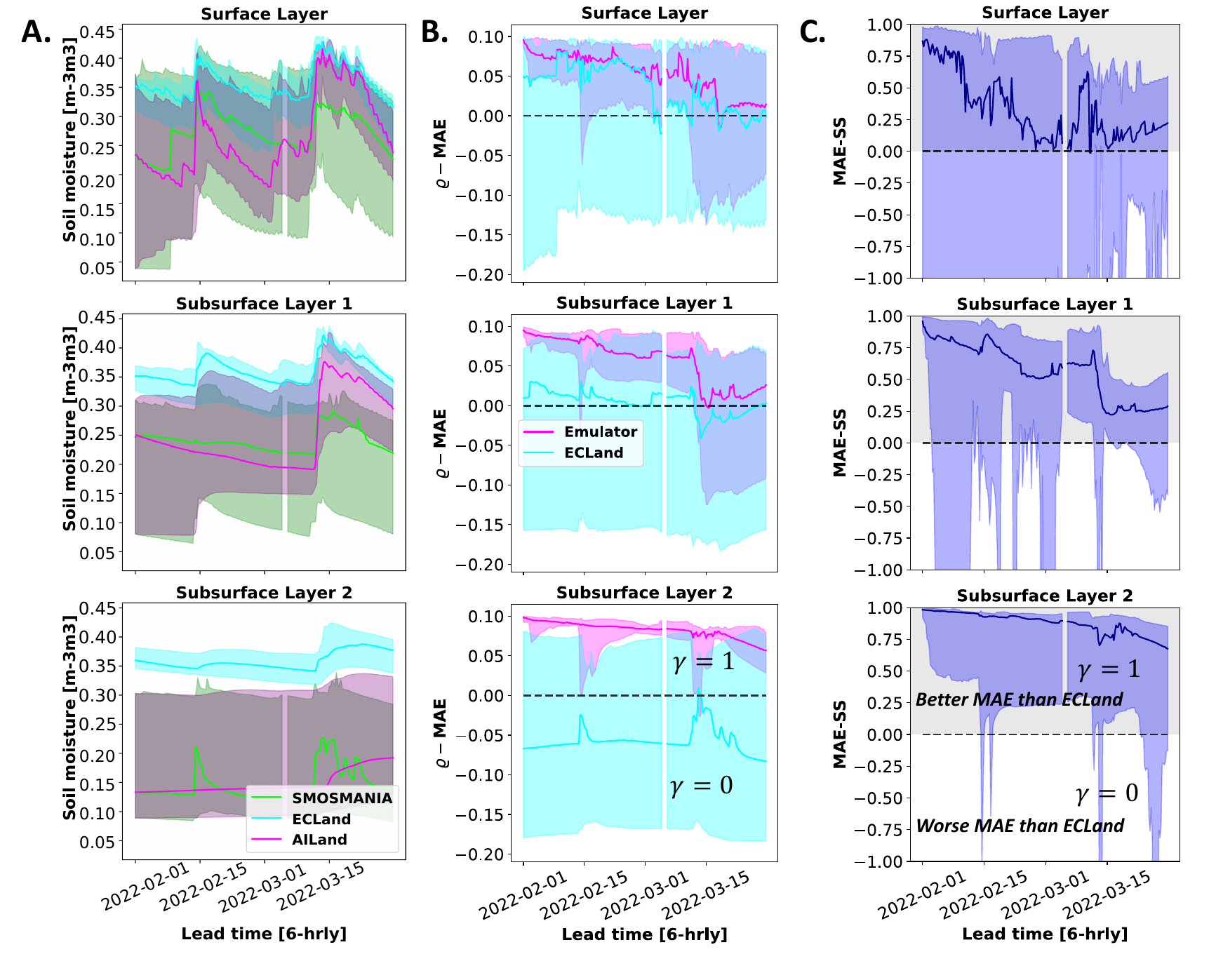}
    \caption{Seasonal-range soil moisture forecasts (A) actual (B) and relative (C) forecast limits, aggregated over eight of 16 stations. 10\% of soil water volume was used as error tolerance. Note however, that the actual amount of water percentage strongly varies by soil type and hence requires an individual threshold per station for full interpretability of the actual forecast limit.}
    \label{fig:aiLand_appendix_moist}
\end{figure}

We found a general agreement in station data and average model forecasts (see figure \ref{fig:aiLand_appendix_temp}, Panel A). Models exhibited a diurnal pattern over the medium-range on the surface layer that attenuated toward deeper layers. At an error tolerance of 1.5 K, ecLand and aiLand show similar and lower average predictability at the surface layer than in the subsurface layers. ecLand has higher spatial variability than aiLand (Panel B). Initialised with observations, aiLand has an advantage over non-initialised ecLand at short lead times (Panel C and figure \ref{fig:aiLand_appendix_temp}): The relative forecast limit of aiLand is 12 hours on the surface layer, nearly two days on subsurface layer 1 and 4 days on subsurface layer 2. After this time period, the information from the observations seems to vanish and aiLand approaches the physical model again.

\begin{figure}[h]
    \centering
    \includegraphics[width=0.99\linewidth]{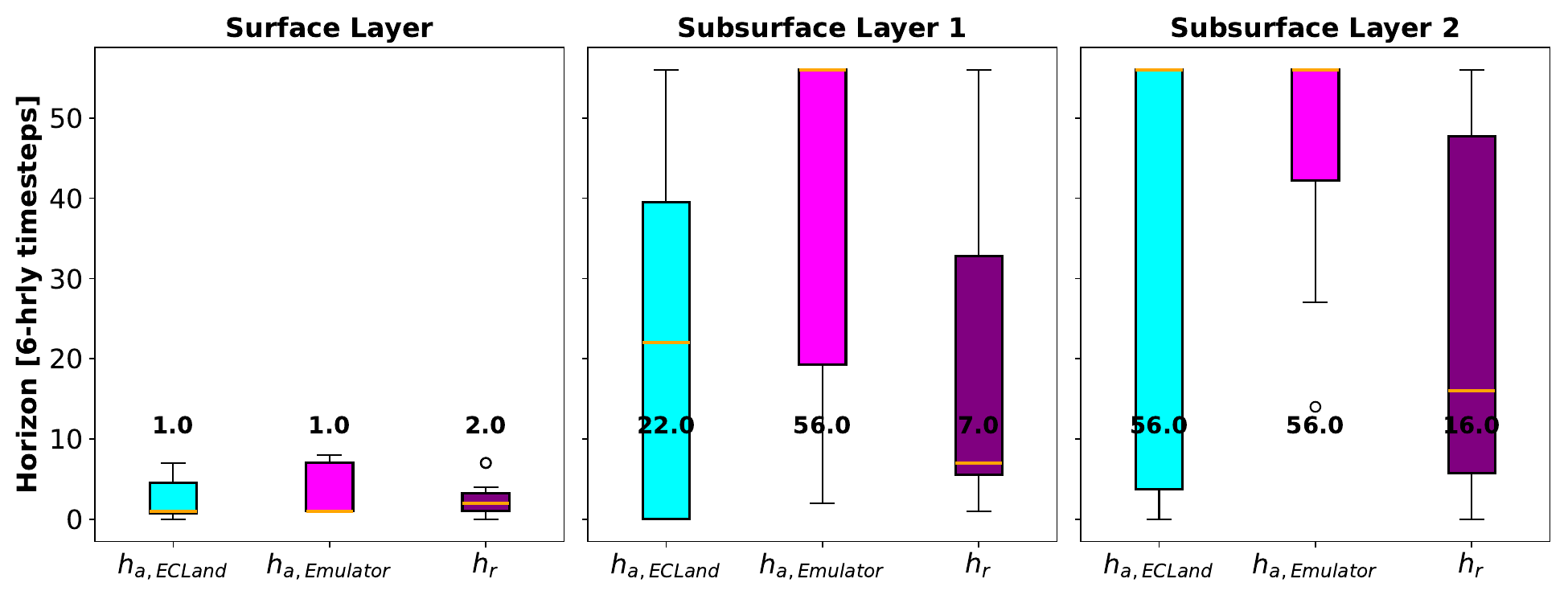}
    \caption{The actual ($h_a$) and relative ($h_r$) empirical forecast limits computed for ecLand and aiLand at three different soil layers. Variability in the box plots refers to variability over network stations. The median limits (orange) indicate an increasing predictability towards lower layers. A positive $h_r$ indicates the limits up to which aiLands initialization with observations shows more skill than ecLand.}
    \label{fig:aiLand_appendix_horizons_temp}
\end{figure}

\begin{figure}
    \centering
    \includegraphics[width=0.95\linewidth]{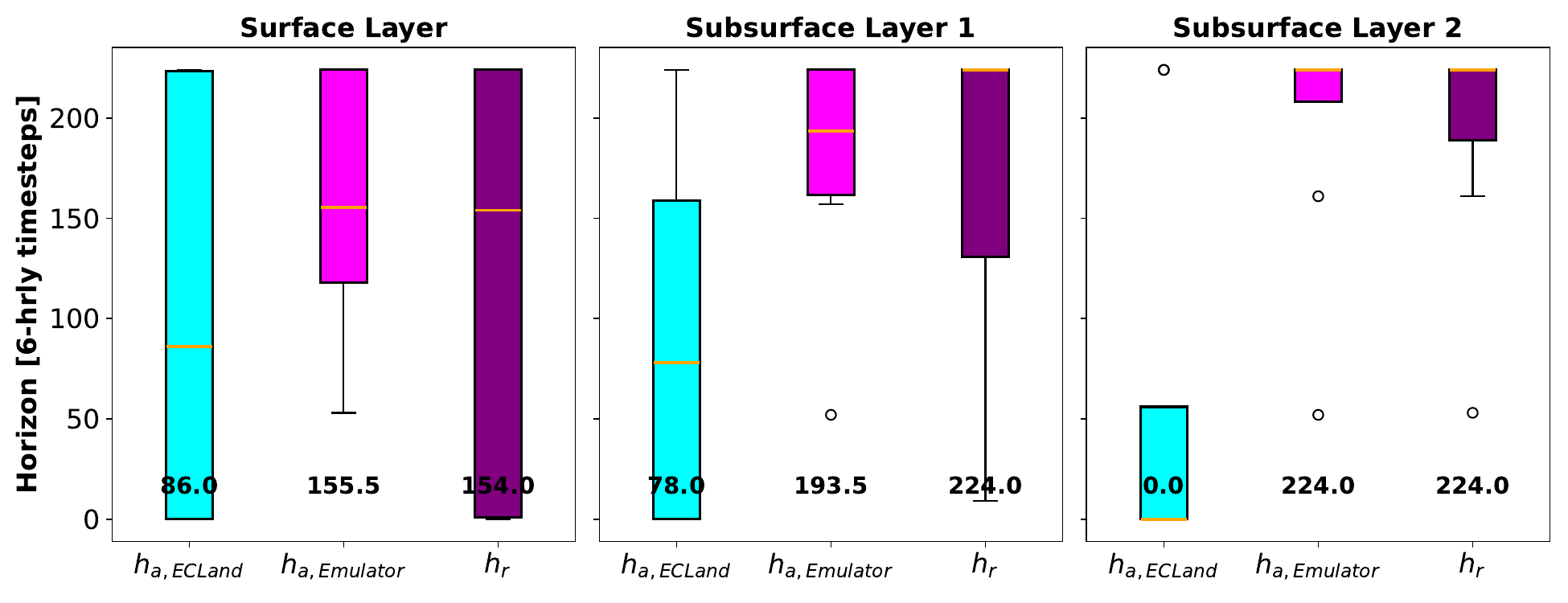}
    \caption{The actual ($h_a$) and relative ($h_r$) empirical forecast limits computed for ecLand and aiLand at three different soil layers. Variability in the box plots refers to variability over network stations. The median limits (orange) indicate an increasing predictability towards lower layers. A positive $h_r$ indicates the limits up to which aiLands initialisation with observations shows more skill than ecLand. Note, that the actual horizon estimate needs to be taken with care, for explanation read discussion paragraph.}
    \label{fig:aiLand_appendix_horizons_moist}
\end{figure}

\paragraph{Discussion} We demonstrate actual and relative forecast limits ($\widehat{h}_a$, $\widehat{h}_r$) of an ensemble of ecLand emulators, aiLand, on the medium-range, testing suitability for weather forecasting. With an scoring tolerance of 1.5 Kelvin we compute $\widehat{h}_a$: the limit up to which aiLand has an acceptable error magnitude for stable modelling of land-atmosphere interactions \citep{boussetta_ecland_2021, zhou_evaluation_2024}. We explore whether the flexibility to quickly initialise prognostic states may be an advantage of aiLand over the physical model. As such, ecLand was not initialised with observations, but aiLand was. The limit up to which this advantage is detectable is indicated by the positive $\widehat{h}_r$ which extends towards lower soil layers. This can be interpreted as an increasing memory for initialisation for slower variables, which is even more visible for soil moisture. The $\widehat{h}_r$ could potentially be further extended by fine-tuning aiLand on station data.
Forecast errors arise not only because of model structures but also due to uncertainty in meteorological forcing: The models forecast to a 31 km spatial grid, while observations are local measurements. As such, when ecLand and aiLand are both driven by erroneous forcing, neither can be expected to have a long $\widehat{h}_a$. 
We show only a local excerpt from aiLands capabilities by looking at limits from a single initialisation time and for soil temperature only. For completeness, we extended the analysis also to the subseasonal range, but without additional results. To better assess predictive ability, the effect of initialisation period on the forecast limits can be explored over the full yearly cycle as in \citet{wesselkamp_advances_2025}. Further, soil temperature interacts strongly with soil moisture and their limits are likely not independent. We did a first analysis on soil moisture limits. However, the same analysis for soil moisture is non-trivial due to its different definitions based on soil type.

\end{document}